\documentclass[11pt, a4paper]{article}
\pdfoutput=1

\usepackage[height=8.85in,width=6.35in]{geometry}

\usepackage{graphicx,rotating}     
\usepackage[bookmarksopen,colorlinks=true,linkcolor=steelblue,
citecolor=darkred,urlcolor=darkred,linktoc=all]{hyperref}

\usepackage{xcolor}
\definecolor{darkred}{rgb}{0.7, 0., 0.}
\definecolor{orangered}{rgb}{1,0.27,0.}
\definecolor{steelblue}{rgb}{0.275,0.51, 0.706}
\definecolor{forestgreen}{rgb}{0.13,0.55,0.13}
\definecolor{violet}{cmyk}{.32,.95,.17,.00}

\usepackage{cite}
\usepackage{amsmath,amssymb,tensor,subdepth}
\usepackage{mathtools}
\usepackage{bm}
\usepackage{bbm}
\usepackage{comment}
\usepackage[utf8]{inputenc}
\usepackage{accents}
\usepackage[footnotesize]{caption}
\usepackage{cancel}
\usepackage{physics}
\usepackage{here}
\usepackage{ascmac}
\usepackage{slashed}
\usepackage{stmaryrd}
\usepackage{trimclip}
\usepackage{tikz-feynman}

\usepackage{chngcntr}
\counterwithin{equation}{section}

\usepackage{libertine}
\usepackage[ISO, upint]{libertinust1math}
\usepackage[scr=boondoxo,bb=boondox]{mathalpha}
\usepackage[T1]{fontenc}

\begin{document}
\hypersetup{pageanchor=false}
\begin{titlepage}

\begin{center}
\hfill TU-1272\\
\hfill KEK-TH-2766\\
\hfill KEK-QUP-2025-0019
\vskip 1.in

\renewcommand{\thefootnote}{\fnsymbol{footnote}}

{\huge \bf
Reheating with Thermal Dissipation and 
\\
Primordial Gravitational Waves\\
}

\vskip .8in

{\Large Kazuma Minami$^{(a)}$,
Kyohei Mukaida$^{(b,c)}$,
Kazunori Nakayama$^{(a,d)}$}

\vskip 0.5in

\begin{tabular}{ll}
$^{(a)}$
& \!\!\!\!\!\emph{Department of Physics, Tohoku University, Sendai 980-8578, Japan}\\[.5em]
$^{(b)}$
& \!\!\!\!\!\emph{Theory Center, IPNS, KEK, 1-1 Oho, Tsukuba, Ibaraki 305-0801, Japan}\\[.5em]
$^{(c)}$
& \!\!\!\!\!\emph{Graduate University for Advanced Studies (Sokendai),}\\
& \!\!\!\!\!\emph{1-1 Oho, Tsukuba, Ibaraki 305-0801, Japan}\\[.5em]
$^{(d)}$
& \!\!\!\!\!\emph{International Center for Quantum-field Measurement Systems for Studies of }\\
& \!\!\!\!\!\emph{the Universe and Particles (QUP), 1-1 Oho, Tsukuba, Ibaraki 305-0801, Japan}
\end{tabular}








\end{center}
\vskip .5in

\begin{abstract}
\noindent
In order for an inflationary universe to evolve into a hot universe, a process known as reheating is required. However, the precise mechanism of reheating remains unknown. We show that if the reheating is triggered by thermal dissipation effects, distinctive features appear in the spectrum of primordial gravitational waves. This suggests a possible way to observationally probe the physics of reheating.

\end{abstract}

\end{titlepage}

\tableofcontents

\renewcommand{\thefootnote}{\arabic{footnote}}
\setcounter{footnote}{0}

\section{Introduction}

During inflationary expansion phase~\cite{Starobinsky:1980te,Guth:1980zm,Sato:1981qmu,Kazanas:1980tx,Linde:1981mu,Albrecht:1982wi}, long wave tensor perturbations of the metric are amplified, which constitute stochastic gravitational wave (GW) background in the present universe~\cite{Starobinsky:1979ty}.
Such a primordial GW background exists in a very wide range of frequencies from cosmological scales to terrestrial scales~\cite{Allen:1987bk,Turner:1990rc,Turner:1993vb,Turner:1996ck,Maggiore:1999vm,Smith:2005mm}.

Since primordial GWs experience all the expansion history of the universe during their propagation, the present-day spectrum of the primordial GW background may have direct information on the early universe.
One of the most important phenomena that must have happened after inflation is the so-called {\it reheating}, in which the energy density stored in the inflaton condensate turns into the radiation energy density. 
It has been shown that the tilt of GW spectrum changes across the frequency corresponding to the comoving Hubble scale at the completion of the reheating, and hence future GW detectors such as DECIGO~\cite{Seto:2001qf} may have a chance to determine the reheating temperature ($T_{\rm R}$) of the universe~\cite{Nakayama:2008ip,Nakayama:2008wy,Kuroyanagi:2008ye,Kuroyanagi:2011fy,Jinno:2014qka}.
The GW spectrum may also contain various information of the early universe such as entropy production~\cite{Seto:2003kc,Nakayama:2009ce,Durrer:2011bi}, phase transition~\cite{Jinno:2011sw}, production of dark radiation~\cite{Jinno:2012xb} through the tensor mode damping effect~\cite{Weinberg:2003ur,Dicus:2005rh,Boyle:2005se}, change of relativistic degrees of freedom~\cite{Watanabe:2006qe,Saikawa:2018rcs}, kination phase~\cite{Tashiro:2003qp,Mukohyama:2009zs} and combinations of all these possible effects in concrete particle physics-motivated models~\cite{Jinno:2013xqa,Ringwald:2020vei}.

In this paper we point out that the GW spectrum contains more detailed information about the reheating.
In the simplest reheating scenario, the inflaton is assumed to decay into lighter species with a perturbative decay rate $\Gamma$, which is just a constant.
However, it is not necessarily true.
Particles in thermal bath may scatter off the inflaton to convert the inflaton energy to the radiation sector: we call this {\it thermal dissipation}~\cite{Yokoyama:2004pf,Yokoyama:2005dv,Drewes:2010pf,Bastero-Gil:2010dgy,Mukaida:2012qn,Mukaida:2012bz,Mukaida:2013xxa,Drewes:2013iaa,Mukaida:2014yia,Mukaida:2014kpa,Tanin:2017bzm,Ai:2021gtg,Ai:2023ahr}.
If the thermal dissipation effect is sizable, the effective decay rate (or the dissipation rate) of the inflaton may have  temperature dependence $\Gamma(T)$.
Although the overall picture that the matter- (or inflaton-) dominated universe connects to the radiation-dominated universe at some temperature $T_{\rm R}$ does not change much qualitatively, the precise evolution around the transition period is affected by the temperature dependence of the dissipation rate.
Therefore, not only the reheating temperature, but also how the reheating proceeds may be imprinted in the GW spectrum.
It will provide a way to probe more detailed properties of the inflaton, such as its mass or interaction strength to the Standard Model particles.

This paper is organized as follows. 
In Sec.~\ref{sec:diss} we briefly summarize thermal dissipation rate by reformulating the derivation of thermal dissipation rate based on the Schwinger--Keldysh effective field theory.\footnote{
    For the sake of completeness, we include inflaton particle production from thermal bath as well since the inflaton particles may alter the reheating process in some cases.
    However, in the following Sec.~\ref{sec:PGW}, we focus on the case where the dissipation of the inflaton condensate complete the reheating for simplicity, and the effect of inflaton particles is neglected.
}
In Sec.~\ref{sec:PGW} we evaluate the primordial GW spectrum under thermal dissipation effect and consider a possibility to distinguish the model of thermal dissipation in future GW experiments.
We conclude in Sec.~\ref{sec:conc}.

\section{Thermal dissipation} \label{sec:diss}
The goal of this section is to provide a formal derivation of the following equation of motion for the inflaton condensate during the stage of reheating:\footnote{
    Although we call $\phi$ the inflaton, it needs not be inflaton itself. It is often the case that some scalar fields other than the inflaton dominates the universe after inflaton decays.
    Here $\phi$ can be any such scalar field with relatively light mass, which finally reheats the universe. Actually thermal dissipation effect is likely to be more relevant for such a scalar field.\label{foot:inf}
}
\begin{align}
	\ddot \phi + \qty[3H + \Gamma (\phi;T) ] \dot \phi + V_\mathrm{eff}'(\phi; T) = 0,
\end{align}
where $H$ is the Hubble parameter and $V_\mathrm{eff}(\phi; T)$ is the thermal effective potential of the inflaton field.
The central question of this section is where the temperature dependence of $\Gamma(\phi;T)$ comes from and how it is calculated.
In the following, we will refomulate Refs.~\cite{Mukaida:2012bz,Mukaida:2012qn,Mukaida:2013xxa} from the viewpoint of Schwinger--Keldysh effective field theory (see \textit{e.g.,} \cite{Liu:2018kfw,Ema:2024hkj}).

\subsection{Preliminaries}

Let $\Gamma_\mathrm{rad}$ being the thermalization rate of radiation.
The precise form of $\Gamma_\mathrm{rad}$ strongly depends on how the radiation is generated from inflaton.
Nevertheless, let us just assume for a while that the thermalization of radiation is much faster than the cosmic expansion
\begin{align}
    \label{eq:thermalplasma}
    H \ll \Gamma_\mathrm{rad}.
\end{align}
One may verify this assumption later by checking the consistency of the resulting equations.
As long as this condition is satisfied, we can treat the radiation as a thermal plasma at a certain temperature $T$.

Suppose that the inflaton field $\phi$ couples to radiation via
\begin{align}
    \label{eq:int_inf-rad}
    \mathcal{L}_\mathrm{int} = \phi O \qty( \{ \chi \} ),
\end{align}
where $O \qty( \{ \chi \} )$ is an operator that depends on fields in the thermal bath, collectively denoted as $\{ \chi \}$.
We do not specify a particular form of $O\qty( \{ \chi \} )$ here; for instance, it could be a Yukawa interaction $\bar \chi \chi$, a trilinear scalar coupling $\chi^2$, or higher dimensional operators such as the gauge kinetic function $F_{\mu\nu} F^{\mu\nu} / \Lambda$ or the Chern--Simons coupling $F_{\mu\nu} \tilde{F}^{\mu\nu} / \Lambda$.
In the following discussion, we mainly consider the case where $\mathcal{L}_\mathrm{int}$ leads to the decay of inflaton at zero temperature.\footnote{
    The case of $\phi^2 \mathcal{O}(\{\chi\})$ type interaction, which does not lead to the decay of inflaton at zero temperature, needs a separate treatment (see \textit{e.g.,} \cite{Mukaida:2013xxa}).
}

\begin{figure}[t]
	\centering
    \includegraphics[width=0.6\linewidth]{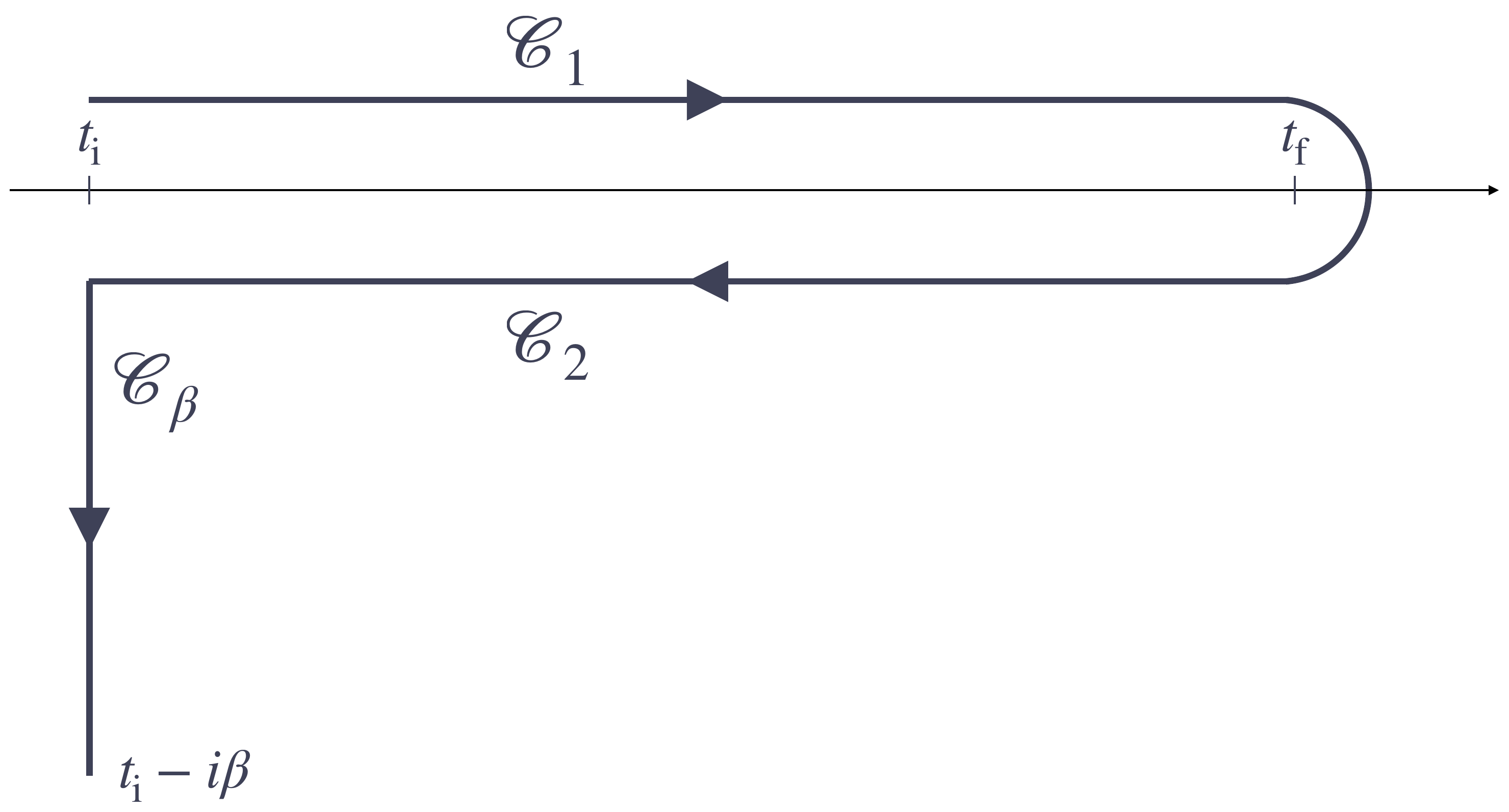}
	\caption{\small 
    The Schwinger--Keldysh contour $\mathcal{C}_{1 + 2 + \beta}$ in the complex time plane is shown. The contour consists of three segments: $\mathcal{C}_1$ (from $t_\mathrm{i}$ to $t_\mathrm{f}$), $\mathcal{C}_2$ (from $t_\mathrm{f}$ to $t_\mathrm{i}$), and $\mathcal{C}_\beta$ (from $t_\mathrm{i}$ to $t_\mathrm{i} - i \beta$).
    $\phi_1$ and $\phi_2$ are defined on $\mathcal{C}_1$ and $\mathcal{C}_2$, respectively.
    If a non-vanishing field value of $\phi$ changes the property of the thermal plasma, $\phi_\mathrm{i}$ resides on the imaginary-time segment $\mathcal{C}_\beta$.
    $\phi_\mathrm{i}$ also lives both on $\mathcal{C}_{1 + 2}$ but these contributions are cancelled out due to the unitarity, and therefore one may only consider $\Delta \phi_{1/2} = \phi_{1/2} - \phi_\mathrm{i}$.
	}
	\label{fig:skcontour}
\end{figure}

Under the condition \eqref{eq:thermalplasma}, one may path-integrate fields in the thermal bath, $\{ \chi \}$, assuming the Gibbs state at temperature $T$.
The inflaton field $\phi$ is treated as a classical background field.
The effective action for the inflaton field is obtained from the following partition function:
\begin{align}
    \label{eq:partfunc}
    e^{i W_T [\phi_1, \phi_2]} &\equiv \Tr \qty[ \rho_T\, \qty{ \tilde{\mathcal{T}} e^{- i \int \dd^4 x\, \mathcal{L}_\mathrm{int} (\phi_2, \{ \chi \}) } } \qty{ \mathcal{T} e^{ i \int \dd^4 x\, \mathcal{L}_\mathrm{int} (\phi_1, \{ \chi \}) } } ] \\
    &=
    \int_{\mathcal{C}_{ 1 + 2 + \beta}} \{\mathcal{D}\chi \} \, e^{i \, \qty( S [\{\chi_1 \}] - S[\{ \chi_2 \}] +  S_\mathrm{int}[\phi_1, \{\chi_1\}] - S_\mathrm{int}[\phi_2, \{\chi_2\}])},
\end{align}
where we have defined
\begin{align}
    S_\mathrm{int}[\phi, \{\chi\}] \equiv \int_{t_\mathrm{i}}^{t_\mathrm{f}} \dd^4 x\, \mathcal{L}_\mathrm{int}(\phi, \{\chi\}).
\end{align}
The time-ordered and anti-time-ordered products are denoted by $\mathcal{T}$ and $\tilde{\mathcal{T}}$, respectively.
The Schwinger--Keldysh contour, $\mathcal{C}_{ 1 + 2 + \beta}$, is a specific path in the complex time plane shown in Fig.~\ref{fig:skcontour}.
The future turning point $t_\mathrm{f}$ is taken to be sufficiently late and the inflaton field satisfies $\phi_1 = \phi_2$ at $t = t_\mathrm{f}$.

Note that one important aspect of the Gibbs state of $\{\chi\}$, denoted as $\rho_T$, is hidden in this expression.
Suppose that the fields $\{\chi\}$ does not involve any conserved quantities.
In this case, one may take the canonical ensemble for $\rho_T$.
The question is whether we should include $\mathcal{L}_\mathrm{int}$ in the definition of the Hamiltonian that defines the canonical ensemble.
If the inflaton condensate does not significantly changes $\rho_T$, one may simply neglect this effect.
However, this is not always the case.
An illustrative examples is when the inflaton condensate changes the effective mass of $\chi$ fields, $m_\chi (\phi)$, for instance, via the Yukawa interaction.
If the effective mass of a certain $\chi$ exceeds the temperature for an initial inflaton field value, \textit{i.e.}, $m_\chi (\phi_\mathrm{i}) \gg T$, this $\chi$ field is not populated in the thermal bath and one may integrate it out from the effective theory.
The resulting interaction of inflaton with the thermal bath is provided by a higher-dimensional operator suppressed by $m_\chi (\phi_\mathrm{i})$.

This observation implies that the canonical ensemble must involve the initial inflaton field value if $\mathcal{L}_\mathrm{int}$ significantly changes the property of the thermal plasma.
To make this effect explicit, we sometimes express the partition function as $W[\Delta \phi_1, \Delta \phi_2; \phi_\mathrm{i}]$ with $\phi = \phi_\mathrm{i} + \Delta \phi$.
If this is the case, the condition \eqref{eq:thermalplasma} alone is not enough to justify the thermalization of $\{\chi\}$ fields along the course of inflaton evolution.
Let $\tau_\phi$ being a time scale when the inflaton field value changes by an order one amount, \textit{i.e.,} $1 \sim \tau_\phi\, \dot \phi_\mathrm{i} / \phi_\mathrm{i}$.
Typically, this time is related with the effective mass of inflaton, $\tau_\phi \sim m_{\mathrm{eff},\phi}^{-1}$, but here we keep a general form.
Now the thermalization condition is expressed as
\begin{align}
    \label{eq:thermalplasma2}
	\tau_\phi^{-1} \ll \Gamma_\text{rad}.
\end{align}
Having this condition in mind, we may assume that the thermalization of $\{\chi\}$ fields is effectively decoupled from the dynamics of the inflaton field, allowing us to treat them as a separate thermal bath.

\subsection{Open effective field theory of inflaton}
Now we are in a position to discuss the effect of the thermal bath imprinted in $W_T[\phi_1, \phi_2]$ on the inflaton dynamics.
We are agnostic about a particular form of $\mathcal{L}_\mathrm{int}$ and write down operators in $W_T[\phi_1, \phi_2]$ allowed by the fundamental symmetries.
As we are integrating out the bath degrees of freedom, the resulting effective theory for the inflaton is open system.
Nevertheless, the partition function is restricted by the unitarity constraints reflecting the fact that the original theory before integrating out the bath is unitary.
The unitarity of the original theory imposes the following three conditions (see \textit{e.g.,} \cite{Liu:2018kfw,Ema:2024hkj}):
\begin{align}
    W_T [\phi, \phi] = 0, \qquad W_T [\phi_1, \phi_2] = - W^\ast [\phi_2, \phi_1], \qquad \operatorname{Im} W_T[\phi_1, \phi_2] \geqslant 0.
\end{align}

On top of this, the nature of the Gibbs state leads to the additional restriction.
Provided that the Gibbs state can be regarded as a complex time evolution along the contour $\mathcal{C}_\beta$, one may show the so-called local Kubo--Martin--Schwinger (KMS) condition~\cite{Glorioso:2016gsa}:\footnote{
    As the inflaton field is dynamical and the resulting $W_T$ is a part of its effective action, we may refer to this condition as dynamical KMS.
    However, the inflaton field can be regarded as an external force for the bath degrees of freedom, and hence this is essentially the local KMS condition for the partition function $W_T$ in the presence of the external force $\phi$.
}
\begin{align}
    W_T [\phi_1, \phi_2] = W_T [\tilde \phi_1, \tilde \phi_2],
\end{align}
where\footnote{
    Here we have assumed that the system has the time reversal symmetry. In general, any discrete symmetry involving the time reversal can be used to derive similar relations, such as $\mathcal{PT}$, $\mathcal{CPT}$ symmetries.
}
\begin{align}
    \tilde \phi_1 (t, {\bm x}) &= \phi_1 (-t + i \beta/2, {\bm x}) = \phi_1 (- t, {\bm x}) - \frac{i\beta}{2} \dot \phi_1 (- t, {\bm x}) + \cdots, \label{eq:cl1}\\
    \tilde \phi_2 (t, {\bm x}) &= \phi_2 (-t - i \beta/2, {\bm x}) = \phi_2 (- t, {\bm x}) + \frac{i\beta}{2} \dot \phi_2 (- t, {\bm x}) + \cdots. \label{eq:cl2}
\end{align}
We have expanded the right-hand sides utilizing the fact that the inflaton fields move more slowly than the thermal degrees of freedom, implying $\beta \ll \tau_\phi$.

Our main interest is the effect of the thermal bath on the dynamics of the inflaton condensate.
To this end, we may consider the classical limit of the partition function $W_T [\phi_1, \phi_2]$.
The classical limit is more transparent in the Keldysh basis, which is defined by
\begin{align}
    \phi_r \equiv \frac{1}{2} \qty(\phi_1 + \phi_2), \qquad \phi_a \equiv \phi_1 - \phi_2.
\end{align}
The $r$-field represents the classical field in the $\hbar \to 0$ limit, while the $a$-field represents the quantum fluctuations of the order $\hbar$ around this classical background.
Indeed, a variation with respect to $\phi_a$ keeping the lowest order in $\phi_a$ gives the classical equation of motion for $\phi_r$.
In the Keldysh basis, the unitarity constraints read
\begin{align}
    W_T [\phi_r, \phi_a = 0] = 0, \qquad W_T [\phi_r, \phi_a] = - W^\ast [\phi_r, -\phi_a], \qquad \operatorname{Im} W_T[\phi_r, \phi_a] \geqslant 0,
\end{align}
and the local KMS is expressed as
\begin{align}
    \label{eq:dkms_r}
    \tilde \phi_r (t, {\bm x}) &= \phi_r (-t, {\bm x}) + \cdots, \\
    \label{eq:dkms_a}
    \tilde \phi_a (t, {\bm x}) &= \phi_a (-t, {\bm x}) - i\beta \dot \phi_r (-t, {\bm x}) + \cdots,
\end{align}
where we have dropped the higher-order terms in $\hbar$ and $\beta/\tau_\phi$.

The classical limit of $W_T [\phi_r, \phi_a]$ consistent with the unitarity and local KMS conditions gives rise to an effective action for the inflaton dynamics in the presence of the thermal bath, which is given by
\begin{align}
    \label{eq:lowenergyeft}
    W_T[\phi_r, \phi_a] = \int \dd^4 x\, \qty(
        i T \Gamma \phi_a^2 - \Gamma \phi_a \Delta \dot \phi_r
        + C_a\, \phi_a + C_{ar}\, \phi_a \Delta \phi_r
    ) + \cdots.
\end{align}
Here the dissipation coefficient $\Gamma$ must be greater than zero, which is originated from the third unitarity constraint.
The non-trivial relation between the coefficients of $\phi_a^2$ and $\phi_a \Delta \phi_r$ is a consequence of the local KMS condition, which is nothing but the fluctuation-dissipation relation.
This property becomes more transparent once we perform the Legendre transformation of $i \,T \Gamma \phi_a^2 \mapsto i \xi^2 / (T \Gamma) + 2 \xi \phi_a$, where $\xi$ is a Gaussian stochastic variable with the variance of $T \Gamma$.
The coefficients $C_a$ and $C_{ar}$ must be real owing to the second unitarity constraints, which may be identified as the Taylor series expansion of the thermal effective potential, \textit{i.e.,}
\begin{align}
    C_a = V_T' (\phi_\mathrm{i}), \qquad C_{ar} = V_T'' (\phi_\mathrm{i}).
\end{align}

The effective action of the inflaton condensate is given by the summation of its treelevel action $S_0$ and the thermal partition function $W_T$ as $\Gamma_\mathrm{eff} \equiv S_0 + W_T$.
The equation of motion for the inflaton condensate is obtained by varying $\Gamma_\mathrm{eff}$ with respect to $\phi_a$ and setting $\phi_a = 0$ afterwards, which yields
\begin{align}
    0 = - \left.\frac{\delta \Gamma_\mathrm{eff}}{\delta \phi_a}\right|_{\phi_a = 0} \simeq \ddot \phi + 3H \dot \phi + V_\mathrm{eff}' + \Gamma \dot \phi,
\end{align}
where we have defined the effective potential as $V_\mathrm{eff}(\phi) \equiv V_0 (\phi) +  V_T(\phi)$ and have utilized $V_T'(\phi) \simeq C_a + C_{ar} \Delta \phi$ with $\phi = \phi_\mathrm{i} + \Delta \phi$.
Note also that one may identify $\phi = \phi_r$ under $\phi_a = 0$.

The same effective action can also describe the dynamics of inflation fluctuations around the homogeneous condensate, corresponding to the \textit{inflaton particles}.
Let the statistical function of $\phi$ being defined as
$
    G_\phi^{rr} (x,y) \equiv \langle \Delta\phi_r (x) \Delta \phi_r (y) \rangle_\text{con}
$, where the subscript ``con'' means that we take only the connected part of the correlation functions.
The equation of motion for $G_\phi^{rr} (x,y)$ is obtained by varying $\Gamma_\mathrm{eff}$, which leads to
\begin{align}
    \left( \partial_t^2 + 3 H \partial_t + \omega_{\bm k}^2 \right) G_\phi^{rr} (t, t'; {\bm k})
    = - \Gamma \partial_t G_\phi^{rr} (t, t'; {\bm k})
    + 2 i T \Gamma G_\phi^{ar} (t, t'; {\bm k}).
\end{align}
Here we have performed the Fourier transformation with respect to ${\bm x} - {\bm y}$ as the system is homogeneous and isotropic, and have defined $\omega_{\bm k}^2 \equiv {\bm k}^2 / a^2 + V_\mathrm{eff}'' (\phi)$ with $a$ being the scale factor of the universe.
Starting from this equation, one may derive the Boltzmann equation for the inflaton particles under a couple of reasonable assumptions (see \textit{e.g.,} \cite{Berges:2004yj}) as follows:
\begin{align}
    \left( \partial_t + 3 H \right) f_\phi (t, {\bm k})
    = - \Gamma \qty[ f_\phi (t, {\bm k}) - \frac{T}{\omega_{\bm k}} ].
\end{align}
One may readily confirm that this coincides with the conventional Boltzmann equation for $\omega_{\bm k} < T$, where we have $f_\phi^\text{(eq)} ({\bm k}) \simeq T / \omega_{\bm k}$.\footnote{
    If one would like to fully recover the Boltzmann equation even valid for $k \sim T$, we cannot take the classical limit [Eqs.~\eqref{eq:cl1} and \eqref{eq:cl2}] and need to keep the non-local structure after integrating out the bath degrees of freedom $\{ \chi \}$.
}
The last term in the right-hand side represents the production of inflaton particles by the thermal bath, which is also controlled by the same dissipation coefficient $\Gamma$ originated from the stochastic noise term.
Therefore, whenever the inflaton condensate dissipates its energy into the thermal bath, the inflaton particles are instead generated from the thermal bath.
For this reason, in general, the dissipation of the inflaton condensate alone is not sufficient to complete the reheating of the universe, and one must verify that the inflaton particles do not dominate the universe after the dissipation of the inflaton condensate.
Indeed, it has been shown that the inflaton particles can re-dominate the universe after the dissipation of the inflaton condensate in some cases~\cite{Mukaida:2013xxa,Fujita:2025zoa,Kaneta:2025xuo}.
See also the discussion at the end of the next subsection.

So far, we have not used the explicit form of $\mathcal{L}_\mathrm{int}$ given in Eq.~\eqref{eq:int_inf-rad}.
Before closing this section, we will provide the matching condition of the effective action \eqref{eq:lowenergyeft} treating the interaction of Eq.~\eqref{eq:int_inf-rad} perturbatively.
The matching can be done by computing the same quantities in the effective action \eqref{eq:lowenergyeft} and the original partition function \eqref{eq:partfunc}, and requiring them to be equal, namely
\begin{align}
    2 T \Gamma \delta (x - y) &= \frac{\delta^2 i W_T}{\delta i \phi_a (x) \delta i \phi_a (y)}\bigg|_{\Delta \phi_r = \phi_a = 0}
    = \left\langle O (\{ \chi_r (x) \}) O (\{ \chi_r (y) \}) \right\rangle_\mathrm{con}
    \equiv G_O^{rr} (x - y), \\
    C_a &= \left. \frac{\delta i W_T}{\delta i \phi_a (x)} \right|_{\Delta \phi_r = \phi_a = 0} = \left\langle O (\{ \chi_r (0) \}) \right\rangle_\mathrm{con}, \\
    C_{ar} \delta (x - y) &= \left. i\, \frac{\delta i W_T}{\delta i \phi_a (x) \delta i \Delta \phi_r (y)} \right|_{\Delta \phi_r = \phi_a = 0} = i \left\langle O (\{ \chi_r (x) \}) O (\{ \chi_a (x) \}) \right\rangle_\mathrm{con} \equiv i\, G_O^{ra} (x - y), \\
    \Gamma \partial_{y_0} \delta (x - y) &= C_{ar} \delta ( x - y ) = i\, G_O^{ra} (x - y).
\end{align}
Here we have also defined the Keldysh basis similarly for $\{\chi\}$ fields as $\chi_r \equiv (\chi_1 + \chi_2)/2$ and $\chi_a \equiv \chi_1 - \chi_2$.
In momentum space, the matching formula takes more familiar form as
\begin{align}
    \label{eq:matching_pspace}
    T \Gamma = \frac{1}{2} \lim_{\omega \to 0} \lim_{{\bm p} \to {\bm 0}} G_O^{rr} (\omega, {\bm p}), \qquad
    C_{ar} = \lim_{\omega \to 0} \lim_{{\bm p} \to {\bm 0}} \operatorname{Re} i\, G_O^{ra} (\omega, {\bm p}),
    \qquad
    \Gamma = \lim_{\omega \to 0} \lim_{{\bm p} \to {\bm 0}} \frac{\operatorname{Im} i\, G_O^{ra} (\omega, {\bm p})}{\omega}.
\end{align}

The first and third matching conditions imply a non-trivial relation between $G_O^{rr}$ and $G_O^{ra}$, which can be understood as a fluctuation-dissipation relation.
One can show this explicitly as follows.
The Kramers--Kronig relation yields
\begin{align}
    \operatorname{Im} i \, G^{ra}_O ( \omega, {\bm p}) = \frac{G_O^\rho (\omega, {\bm p})}{2},
    \qquad
    G_O^\rho (x - y) \equiv \left\langle \left[ O (\{ \chi_r (x) \}), O (\{ \chi_r (y) \}) \right] \right\rangle,
\end{align}
On the other hand, the Kubo--Martin--Schwinger (KMS) relation leads to
\begin{align}
    G_O^{rr} ( \omega, {\bm p}) = \qty[ \frac{1}{2} + f_\mathrm{BE} ( \omega ) ] G_O^{\rho} ( \omega, {\bm p}).
\end{align}
Combining these two relations, one readily finds
\begin{align}
    \frac{1}{2 T} \lim_{\omega \to 0} \lim_{{\bm p} \to {\bm 0}} G_O^{rr} (\omega, {\bm p})
    =
    \lim_{\omega \to 0} \lim_{{\bm p} \to {\bm 0}} \frac{G_O^{ra} (\omega, {\bm p})}{2 \omega}
    = \lim_{\omega \to 0} \lim_{{\bm p} \to {\bm 0}} \frac{\operatorname{Im} i\, G_O^{ra} (\omega, {\bm p})}{\omega},
\end{align}
which is consistent with the first and third matching conditions in Eq.~\eqref{eq:matching_pspace}.

\subsection{Examples} \label{sec:example}

\paragraph{Scalar trilinear interaction.}
Let us start with a simple example of
\begin{align}
    \mathcal{L}_\mathrm{int} = - A\, m_\phi \phi\, \abs{\chi}^2,
\end{align}
where $\chi$ is a complex scalar field in the thermal bath charged under a certain gauge symmetry, $m_\phi$ is the inflaton mass at the potential minimum, and $A$ is a dimensionless coupling constant.
The thermal mass of $\chi$ is estimated as $m_{\chi T}\sim g T$ with $g$ being a gauge coupling constant.

In the following discussion, we consider the regime where the tachyonic preheating is inefficient, which implies $A \tilde \phi \ll m_\phi$ with $\tilde \phi$ being the amplitude of the inflaton oscillation.
Then, for $m_\phi \gg m_{\chi T}$, the dissipation rate of the inflaton is just given by the perturbative decay rate at zero temperature:
\begin{align}
    \Gamma \sim A^2 m_\phi.
\end{align}
On the other hand, if the inflaton oscillation is much slower than the thermalization rate $m_\phi \ll \Gamma_\mathrm{rad}$, the dissipation rate is given by \cite{Mukaida:2012qn,Mukaida:2012bz}
\begin{align}
    \Gamma = \lim_{\omega \to 0} \frac{G^\rho_{|\chi|^2} (\omega, {\bm 0})}{2\omega} 
    \sim \frac{ A^2  m_\phi^2}{\gamma_\chi}
    \propto T^{-1}.
\end{align}
Here $\gamma_\chi$ is the scattering rate of $\chi$ relevant for the relaxation of $|\chi|^2$ perturbation, which is proportional to $T$ for relativistic particles.

\paragraph{Yukawa interaction.}
Next we consider the Yukawa interaction
\begin{align}
    \mathcal{L}_\mathrm{int} = - y \phi \bar \psi \psi,
\end{align}
where $\psi$ is a Dirac fermion in the thermal bath charged under a certain gauge symmetry, and $y$ is a dimensionless coupling constant.
The thermal mass of $\psi$ is estimated as $m_{\psi T} \sim g T$ with $g$ being a gauge coupling constant.

Similarly to the previous example, for $y \tilde \phi \ll m_\phi$ and $m_\phi \gg m_{\psi T}$, the dissipation rate is given by the perturbative decay rate at zero temperature
\begin{align}
    \Gamma \sim y^2 m_\phi,
\end{align}
while for $m_\phi \ll \Gamma_\mathrm{rad}$, the dissipation rate is given by~\cite{Mukaida:2012qn,Mukaida:2012bz}
\begin{align}
    \Gamma = \lim_{\omega \to 0} \frac{G^\rho_{\bar \psi \psi} (\omega, {\bm 0})}{2\omega} 
    \sim y^2 \alpha T \propto T.
\end{align}
The latter one stands for the scattering involving the gauge boson, \textit{e.g.,} $\phi \chi \to g \chi$.
For a slightly larger inflaton amplitude, $\Gamma_\text{rad} \ll y\tilde \phi \ll T$, the Dirac mass term from the inflaton condensate becomes important and the dissipation rate becomes~\cite{Mukaida:2012qn,Mukaida:2012bz}
\begin{align}
    \Gamma \sim \frac{y^4 \tilde \phi^2}{\gamma_\psi} \propto  \frac{\tilde \phi^2}{T}.
\end{align}
Again $\gamma_\psi$ is the relevant scattering rate of $\psi$, which is proportional to $T$ for relativistic particles.

If the inflaton amplitude is further increased to $y \tilde \phi \gg T$, one may integrate out $\psi$ from the effective theory, which leads to the following interaction~\cite{Mukaida:2012qn,Mukaida:2012bz}:
\begin{align}
    \mathcal{L}_\mathrm{int} = - \frac{\Delta \phi}{\tilde \phi} \frac{\alpha}{4 \pi} \frac{2 \operatorname{T}_\psi}{3} \, F^a_{\mu\nu} F^{a \mu\nu} + \cdots,
\end{align}
where $\operatorname{T}_\psi$ is the normalization of generators associated with the representation of $\psi$, and the fine structure constant is denoted by $\alpha = g^2 / (4 \pi)$.
The dissipation rate is then given by
\begin{align}
    \Gamma = \frac{4 \operatorname{T}^2_\psi}{9} \frac{\alpha^2}{16 \pi^2 \tilde \phi^2} \lim_{\omega \to 0} \frac{G^\rho_{FF} (\omega, {\bm 0})}{2\omega} \sim \frac{\alpha^4 T^4}{\tilde \phi^2 \gamma_g } \propto \frac{T^3}{\tilde \phi^2},
\end{align}
where $\gamma_g$ is the relevant interaction rate of gauge bosons proportional to $T$ for relativistic particles.
In the first similarity, we have used the parametric estimate of $FF \sim g^2 T^2 A A$.

If the inflaton condensate oscillates around a non-zero VEV, the above interactions become less efficient later as it decreases faster than the Hubble parameter.
In this case, the inflaton particles produced from the thermal bath may dominate the universe after the dissipation of the condensate, which requires a more careful analysis \cite{Fujita:2025zoa}.

\section{Effects on primordial gravitational waves} \label{sec:PGW}

\subsection{Primordial gravitational waves}

The tensor perturbation of the metric around the Friedmann-Robertson-Walker metric is written as
\begin{align}
	\dd s^2=-\dd t^2 + a^2(t)(\delta_{ij} + h_{ij}(t,\vec x)) \dd x^i \dd x^j.
\end{align}
It must satisfy the transverse-traceless condition $\partial_i h_{ij} = h_{ii} = 0$ and this is regarded as GW. It is expanded as
\begin{align}
	h_{ij}(t,\vec x) = \sum_{\lambda=+,\times} \int\frac{\dd^3k}{(2\pi)^3} \left[ a_{k,\lambda} h_{k}^{\lambda}(t) + a^\dagger_{-k,\lambda} h^{\lambda *}_{k}(t)  \right]e^{i\vec k\cdot\vec x} e^\lambda_{ij},
\end{align}
where the polarization tensor satisfies $e_{ij}^\lambda e_{ij}^{*\lambda'} = \delta_{\lambda\lambda'}$ and the creation-annihilation operator satisfies the following commutation relation:
\begin{align}
	[a_{k,\lambda}, a^\dagger_{k',\lambda'}] = (2\pi)^3 \delta_{\lambda\lambda'}\delta(\vec k-\vec k').
\end{align}
The equation of motion of the Fourier mode is given by
\begin{align}
	\ddot h^\lambda_{k} + 3H \dot h^\lambda_k + \frac{k^2}{a^2} h^\lambda_k = 0. \label{eom_h}
\end{align}

Let us define several quantities for later convenience.
The GW power spectrum $P_h(t,k)$ is defined as
\begin{align}
	\left<h^2_{ij} (t)\right> = 2 \int\frac{\dd^3k}{(2\pi)^3}\left| h^\lambda_k(t) \right|^2 = \int\frac{\dd^3k}{(2\pi)^3} P_h(k,t)
	= \int \dd \ln k\,\Delta_h^2(t,k)
	= 2 \int \dd f S_h(f),
\end{align}
where the factor 2 comes from two polarizations and we have also defined the spectral density $S_h(f)$ with $f=k/(2\pi a(t))$ and the dimensionless power spectrum as
\begin{align}
	\Delta^2_h(t,k) \equiv \frac{k^3}{2\pi^2} P_h(t,k) = \frac{k^3}{\pi^2} \left| h^\lambda_k(t) \right|^2.
\end{align}
Its initial condition is set by the superhorizon tensor perturbation generated during inflation:
\begin{align}
	\Delta^2_h(t\to 0,k) = 8 \left(\frac{H_{\rm inf}(k_*)}{2\pi M_{\rm Pl}}\right)^2\left(\frac{k}{k_*}\right)^{n_t} \equiv \left[\Delta^{\rm (ini)}_h(k)\right]^2,
	\label{Delta_inf}
\end{align}
where $H_{\rm inf}(k_*)$ denotes the Hubble scale during inflation when the comoving wave number $k_*$ exits the horizon, $M_{\rm Pl}$ is the reduced Planck scale, and $n_t$ the tensor spectral index. 
In standard slow-roll inflation models, the relation $n_t = -2\epsilon$ holds with $\epsilon$ being the slow-roll parameter. Thus $n_t$ is small enough and hence GWs have nearly scale-invariant spectrum. In the numerical study below, we take $n_t=0$ for simplicity.
The energy density of the GW is given by
\begin{align}
	\rho_h(t,\vec x) = \frac{M^2_{\rm Pl}}{4} \left<\dot h_{ij} \dot h_{ij} \right> \equiv \int \dd\ln k\,\rho_h(t, k),
\end{align}
where
\begin{align}
	\rho_h(t, k) = \frac{M_{\rm Pl}^2}{2}\frac{k^3}{2\pi^2} \left\langle \left|\dot h^\lambda_k(t)\right|^2 \right\rangle_{\rm osc}
	= \frac{M_{\rm Pl}^2}{4} \frac{k^3}{2\pi^2}\left( \left| \dot h^\lambda_k(t) \right|^2 + \frac{k^2}{a^2}\left| h^\lambda_k(t) \right|^2 \right),
\end{align}
for $k\gg aH$ by taking the oscillaton average.
The present-day GW spectrum is represented in terms of the GW density parameter $\Omega_h(f)$, defined as
\begin{align}
	\Omega_h(f) = \frac{\rho_h(t_0,f)}{\rho_{\rm crit}(t_0)} = \frac{2\pi^2 f^3}{3H_0^2} S_h(f),
\end{align}
where $t_0$ is the present age of the universe, $\rho_{\rm crit}$ is the critical energy density and $H_0$ is the present Hubble parameter.
The comoving wave number of the GW $k$ is converted to the present frequency $f$ through $f = k/(2\pi a(t_0))$.

\subsection{Time evolution of gravitational waves}

The solution of the equation of motion (\ref{eom_h}) is given by
\begin{align}
	h_k \propto \begin{cases}
		{\rm const.} & {\rm for}~k \lesssim aH \\
		1/a(t) &  {\rm for}~k \gtrsim aH 
	\end{cases}.
\end{align}
Thus the energy density of GW decreases as $\rho_h(t,k)\propto a^{-4}$ inside the horizon, as expected.
It is not hard to show that the resulting GW energy spectrum at some fixed time $t$ looks like $\rho_h(t,k) \propto k^0$ ($\rho_h(t,k) \propto k^{-2}$) for the modes which entered the horizon in the radiation- (matter-) dominated era~\cite{Nakayama:2008ip,Nakayama:2008wy}. 
Therefore, the present-day GW spectrum contains information of thermal history of the universe. 
In particular, the reheating temperature $T_{\rm R}$ can be directly extracted from the frequency at which the tilt of the GW spectrum changes:
\begin{align}
    f_{\rm R} = \frac{(aH)_{3H=\Gamma_0}}{2\pi a(t_0)} \simeq 0.26\,{\rm Hz}\left(\frac{T_{\rm R}}{10^7\,{\rm GeV}}\right)\left(\frac{g_{*s}(T_{\rm R})}{106.75}\right)^{1/6},
\end{align}
where $g_{*s}$ is the effective relativistic degrees of freedom for the entropy density.

\begin{figure}
    \centering
    \includegraphics[width=.45\textwidth ]{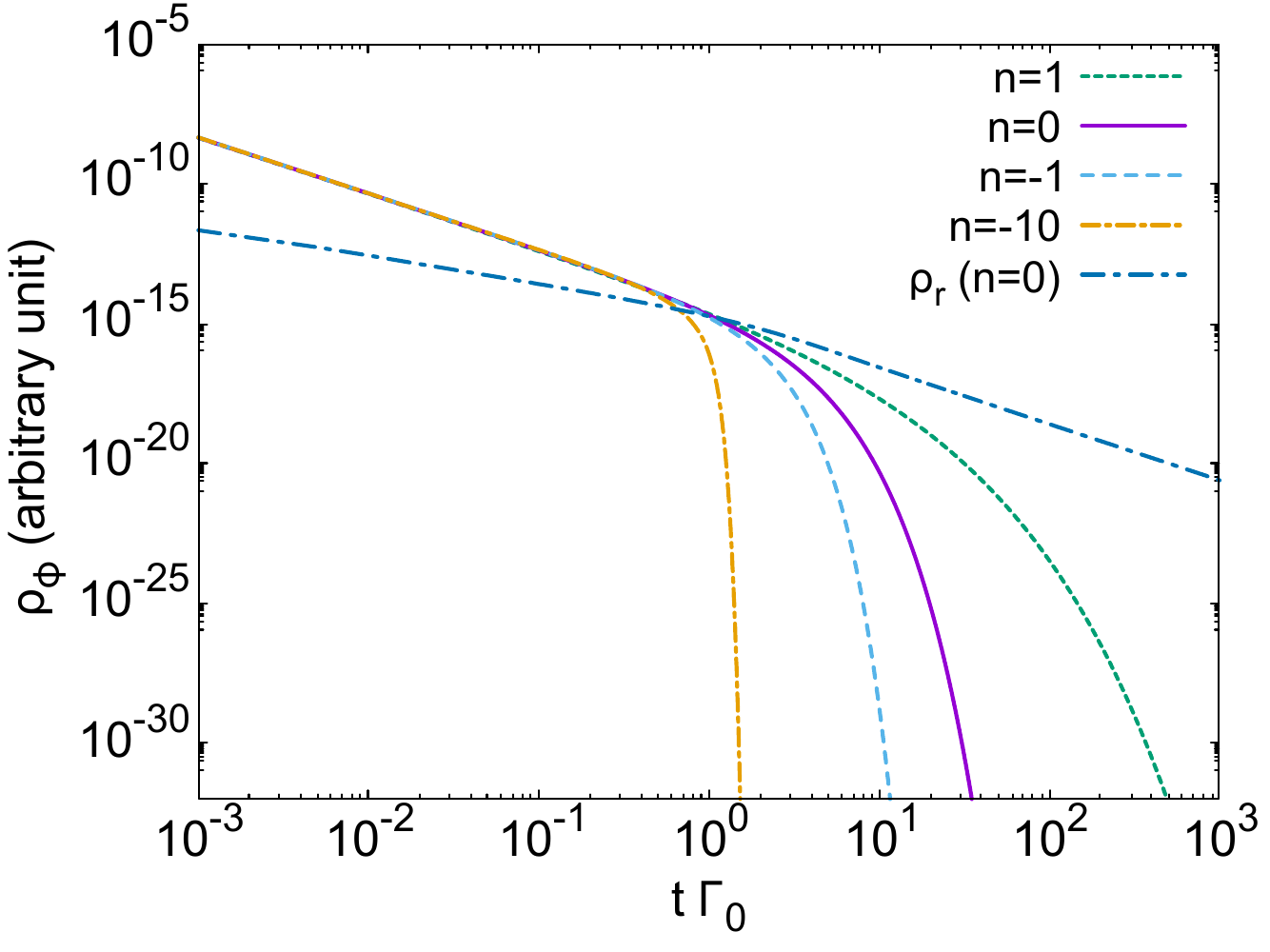}
    \includegraphics[width=.45\textwidth ]{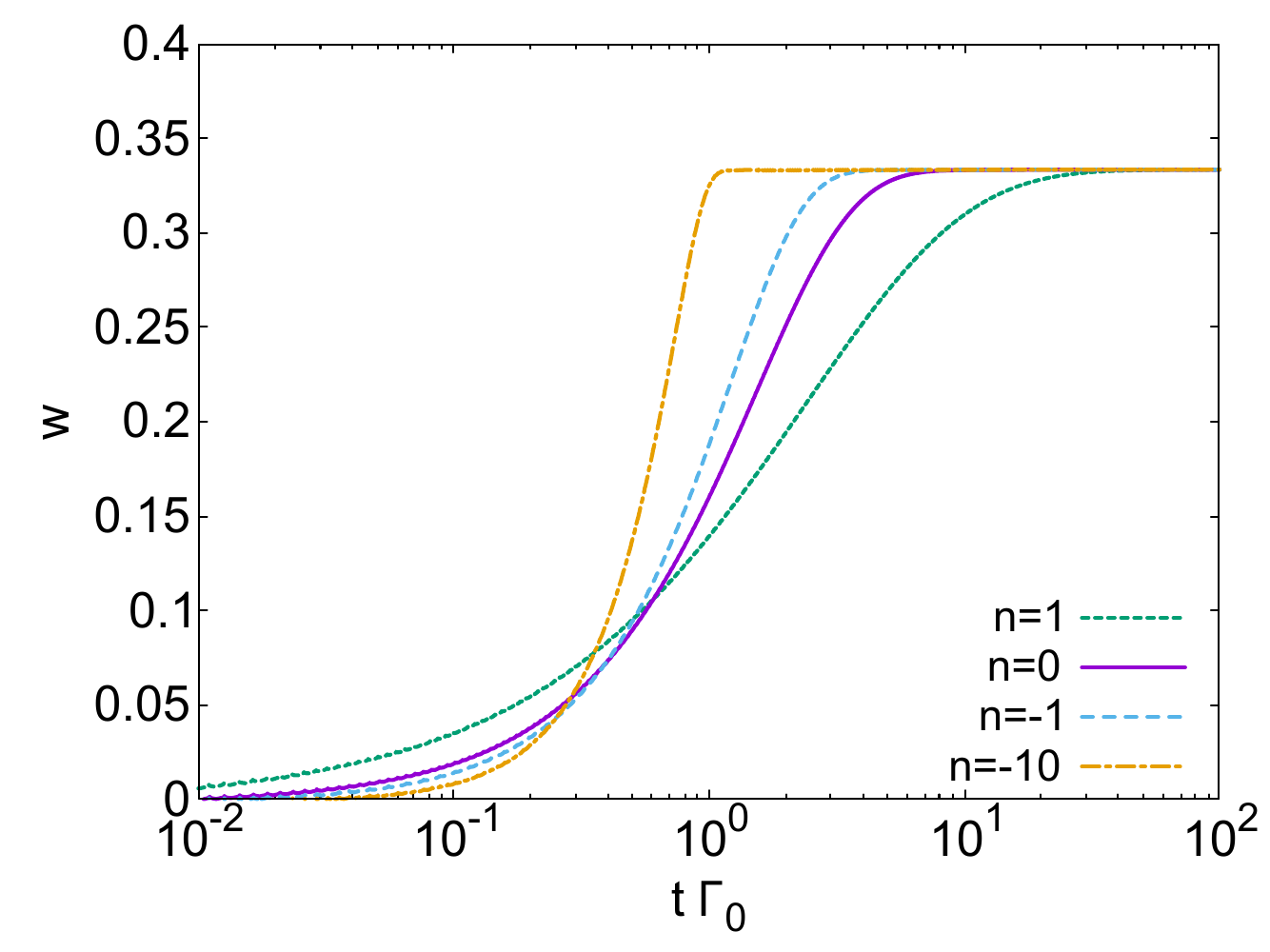}
    \caption{(Left) Time evolution of the inflaton energy density $\rho_\phi$ for different thermal dissipation model $n=1, 0, -1$ and $-10$ in (\ref{GammaT}). For comparison, $\rho_r$ for $n=0$ is also plotted. (Right) Time evolution of the equation of state parameter $w$ for $n=1, 0, -1$ and $-10$.}
    \label{fig:w}
\end{figure}

Now we will more closely discuss the imprints of reheating on the GW spectrum. 
To precisely calculate the GW spectrum, we must solve the set of equations
\begin{align}
	&3M_{\rm Pl}^2 H^2 = \rho_\phi + \rho_r, \\
	&\dot\rho_\phi + 3H \rho_\phi = -\Gamma(T) \rho_\phi,\\
	&\dot\rho_r + 4H \rho_r = \Gamma(T) \rho_\phi,
\end{align}
along with the GW equation of motion (\ref{eom_h}), where $\rho_\phi$ and $\rho_r$ are the energy density of the inflaton and radiation, respectively.\footnote{
    Again we note that $\phi$ needs not be the inflaton. See footnote \ref{foot:inf}.
}
The radiation energy density is expressed in terms of the temperature $T$ as $\rho_r = \pi^2 g_* T^4/30$ with $g_*$ being the relativistic degrees of freedom. The decay rate of the inflaton, taking account of the thermal dissipation effect, is represented by $\Gamma(T)$ as a function of temperature.
Below we phenomenologically parametrize the decay rate as
\begin{align}
	\Gamma(T) = \Gamma_0 \left(\frac{T}{T_{\rm R}}\right)^n,\qquad
    \Gamma_0 \equiv \left(\frac{\pi^2 g_*}{10}\right)^{1/2}\frac{T_{\rm R}^2}{M_{\rm Pl}}.\label{GammaT}
\end{align}
The standard perturbative decay corresponds to $n=0$.
As reviewed in Sec.~\ref{sec:diss}, the value of $n$ depends on the types of the interaction.
Typical examples examined in Sec.~\ref{sec:example} correspond to $n=1, 3$ or $-1$, but in the following we regard $n$ as a phenomenological parameter. 
Note that, in order for $\Gamma(T)$ to eventually overcome $H$ so that the reheating will be completed, we need $n<2$.\footnote{
	We can show this as follows. The radiation density produced per Hubble time is given by $\rho_r \sim \rho_\phi \Gamma(T)/H \propto H T^n$. It gives $T \propto H^{1/(4-n)}$ and hence $\Gamma(T) \propto H^{n/(4-n)}$. We need $n<5/2$ for consistency of our scenario (see footnote~\ref{foot:preexist}). For completion of the reheating, $\Gamma(T)$ must decrease more slowly than $H$. Thus we need $n/(4-n) < 1$ or $n<2$.
}
As far as this condition is satisfied, the overall picture of thermal history does not much depend on $n$: the inflaton-dominated matter phase ends around $H=\Gamma_0$ and the radiation-dominated phase starts with $T=T_{\rm R}$.  
Still, the precise background evolution depends on $n$ at the transition epoch around $T=T_{\rm R}$. 
For $n=1$, for example, the transition is flatter than the standard case $n=0$, while for negative $n$ the transition becomes sharp.
Fig.~\ref{fig:w} shows the time evolution of the inflaton energy density and the equation of state parameter $w$ for $n=1, 0, -1$ and $-10$, where we have numerically calculated $w$ through
\begin{align}
	w = -1 - \frac{1}{3} \frac{\dd \ln \rho_{\rm tot}}{\dd \ln a},\qquad \rho_{\rm tot} = \rho_\phi + \rho_r.
\end{align}
It is clearly seen that the transition becomes sharper for smaller $n$.
This difference of $w$ around the transition period may be imprinted in the GW spectrum.

Note that, for $n\neq 0$, we need to assume that initially radiation is produced by some mechanisms other than thermal dissipation, such as perturbative decay or preheating.\footnote{
    The energy density of pre-existing radiation decreases as $\rho_r^{\rm (pre)} \propto a^{-4}$, while the newly produced radiation through the dissipation effect scales as $\rho_r^{\rm (diss)} \propto H T^n$. Initially $\rho_r^{\rm (pre)} > \rho_r^{\rm (diss)}$ and hence $T\propto a^{-1}$, implying $\rho_r^{\rm (diss)} \propto a^{-(n+3/2)}$. Thus we need $n<5/2$ in order for the newly produced radiation eventually overcomes the pre-existing radiation.
    \label{foot:preexist}
} Actually the effective dissipation rate (\ref{GammaT}) does not continue to take the identical form, but changes its form in realistic models. However, our discussion remains intact as far as the dissipation rate takes the form of (\ref{GammaT}) only around the completion of reheating.

\begin{figure}
    \centering
    \includegraphics[width=.45\textwidth ]{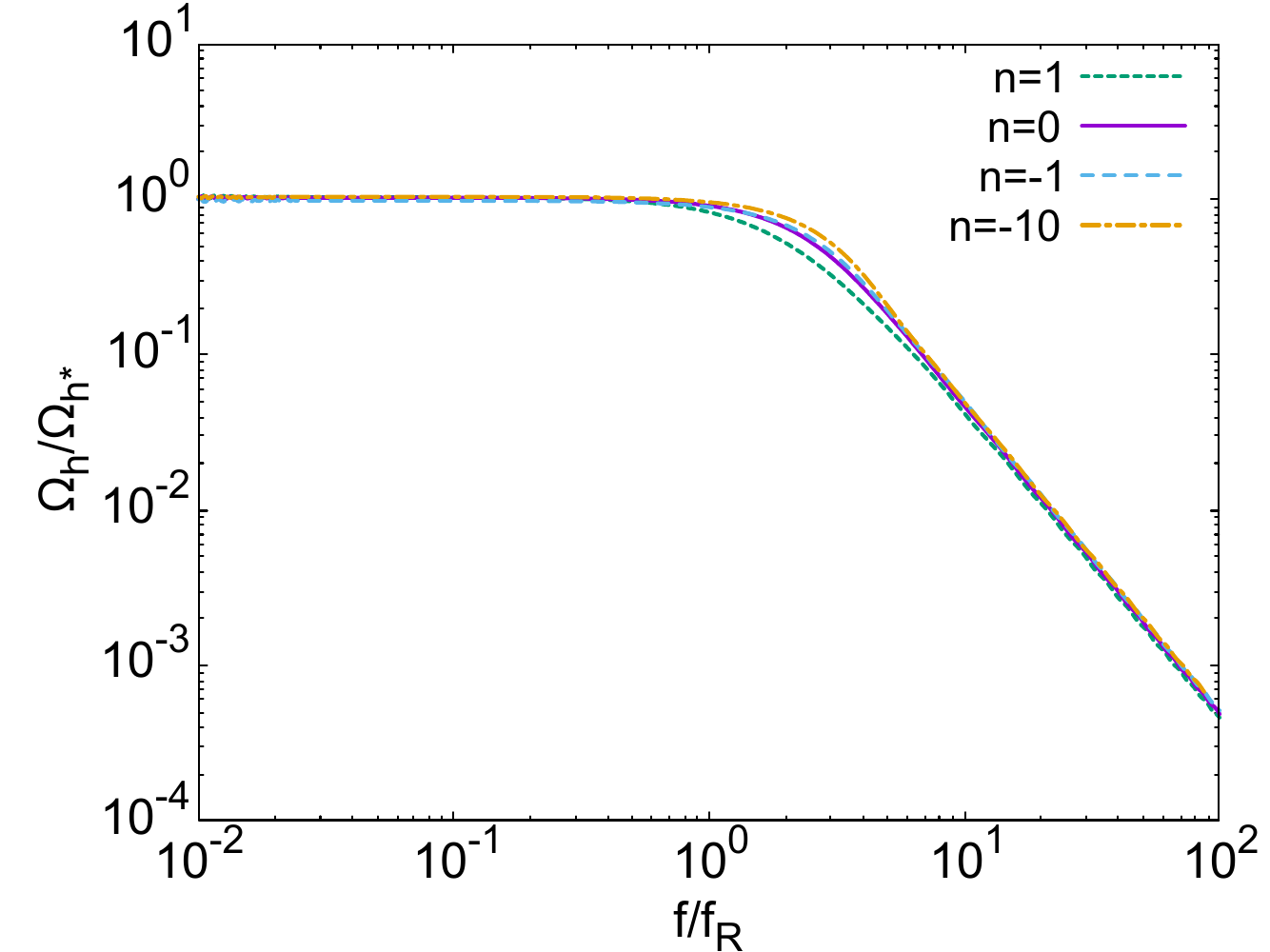}
    \hspace{5mm}
    \includegraphics[width=.45\textwidth ]{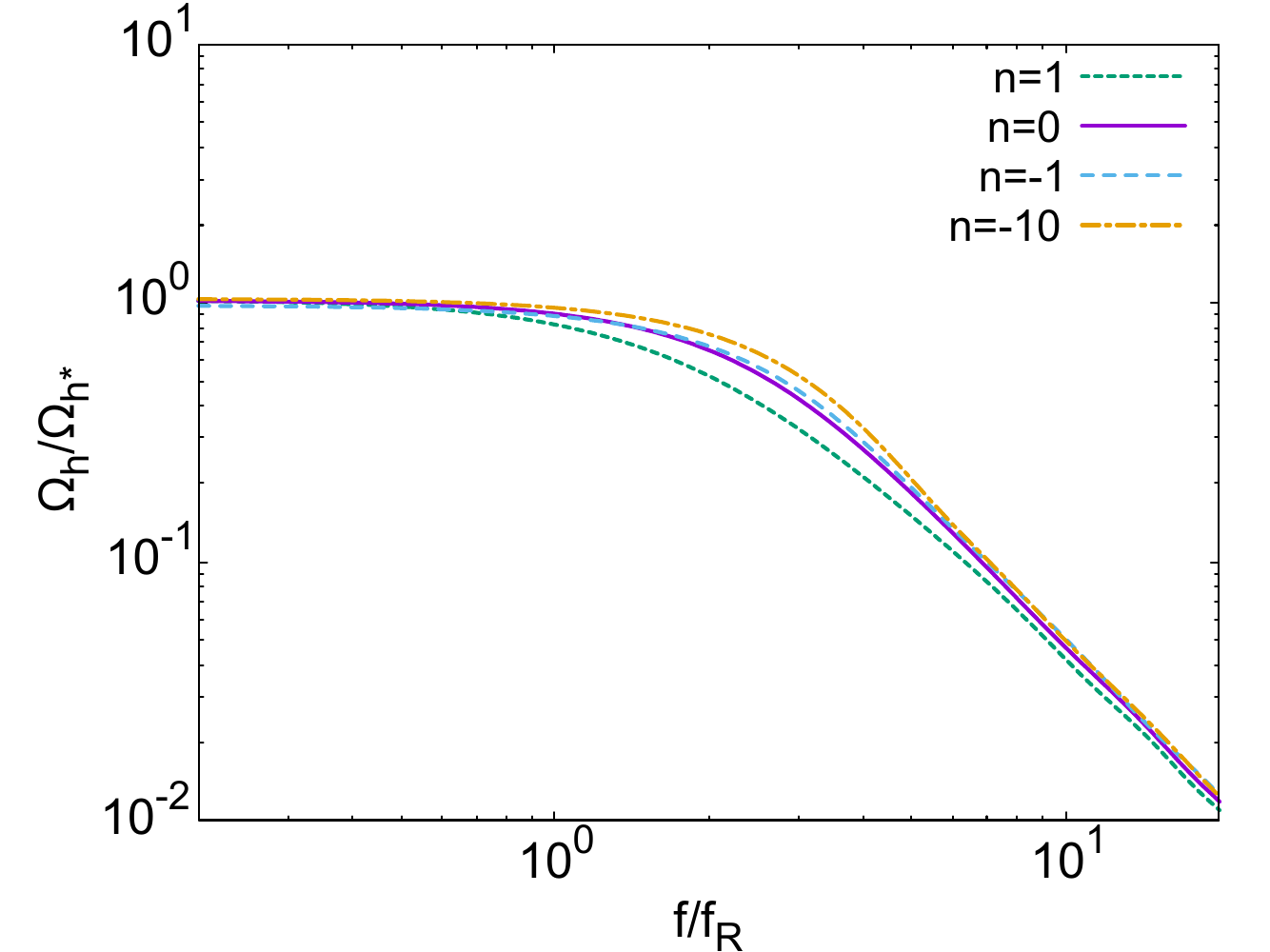}
    \caption{(Left) GW spectrum with reheating under thermal dissipation (\ref{GammaT}). The case of $n=1,0,-1$ and $-10$ are shown. They are normalized so that the leftmost part becomes equal to unity. (Right) Enlarged view of the left figure.}
    \label{fig:Oh}
\end{figure}

The result for the GW spectrum is shown in Fig.~\ref{fig:Oh}.
We have shown the results for $n=1,-1$ and $-10$ in comparison with the $n=0$ case. 
The horizontal axis is normalized by the transition frequency $f_{\rm R}$.\footnote{
    Here is a technical remark. For the purpose of comparing the GW spectra with different $n$, we need to match both the low- and high-frequency ends of the spectra. In order to do so, $\Gamma_0$ (or $T_{\rm R}$) should be slightly shifted (about 10-20 percent) for different $n$. Our presented value of $T_{\rm R}$ in Sec.~\ref{sec:imp} corresponds to that for $n=0$.
}
It is seen that the case of $n=1$ clearly deviates from $n=0$ around the transition frequency.
On the other hand, the difference between the $n=-1$ and $n=0$ cases is hardly seen by eye. 
The difference becomes clearer for negatively large $n$ like $n=-10$ as shown in the figure. We have numerically checked that further decreasing $n$ does not create much more difference. 
The reason why there does not appear big difference for negatively large $n$ may be understood as follows.
For negatively large $n$, the transition from the matter- to radiation-dominated phase becomes sharper.
However, the GW wavelength corresponding to the transition epoch is around the Hubble scale at that epoch. Thus it does not  ``feel'' the rapid change of the equation of state.
 On the other hand, it is sensitive to the slower change of the equation of state, which is the case for $n=1$.

\subsection{Implications for future observations}
\label{sec:imp}

We have seen that the thermal dissipation effect may be imprinted in the GW spectrum.
Now we discuss possibility for distinguishing the power $n$ in future GW observations, in particular DECIGO experiment~\cite{Seto:2001qf}.

To find stochastic GW signals, we take a correlation between two detectors. The signal-to-noise ratio (SNR) is expressed as~\cite{Allen:1997ad,Maggiore:1999vm,Maggiore:2007ulw}
\begin{align}
	{\rm SNR}(f) = \sqrt{2t_{\rm obs} \Delta f} \frac{F \Omega_h(f)}{\Omega_n(f)},  \label{SNR}
\end{align}
where $t_{\rm obs}$ is the total observation time, $\Delta f$ is the frequency bin, which is taken arbitrary to maximize the desired signal, $F$ is $\mathcal O(1)$ constant representing the detector response to signal GWs, and $\Omega_n(f)$ is the noise spectrum, usually consisting of shot noise, radiation pressure noise, and acceleration noise.\footnote{
	The expression (\ref{SNR}) assumes that the signal and noise spectrum can be approximated  as constant within the frequency bin $\Delta f$. It is  valid for our purpose.  
}
These details are summarized in App.~\ref{sec:noise}.
We use the typical DECIGO noise spectrum $\Omega_n(f)$ with Fabry-Perot (FP) interferometry (FP-DECIGO).
Then we simply regard $F^{-1} \Omega_n(f) / \sqrt{2t_{\rm obs} \Delta f}$ as an expected error on the measured $\Omega_h(f)$.
For comparison, we also adopt the noise spectrum of ultimate-DECIGO~\cite{Kudoh:2005as,Kuroyanagi:2014qza} to show the ultimate possibility to probe the details of the reheating.

Before performing a numerical analysis, let us make a rough estimation.
The GW spectrum at the flat part $(f \ll f_{\rm R})$ is evaluated as
\begin{align}
	\Omega_h(f)\simeq 1.2\times 10^{-15} \left(\frac{H_{\rm inf}(k_*)}{10^{14}\,{\rm GeV}}\right)^2
        \left(\frac{g_*(T)}{g_*(T_0)}\right) \left(\frac{g_{*s}(T_0)}{g_{*s}(T)}\right)^{4/3}.
\end{align}
Comparing it with the sensitivity of FP-DECIGO, we need $H_{\rm inf} \gtrsim {\rm (a~few)}\times 10^{13}\,{\rm GeV}$ and also $T_{\rm R} \gtrsim ({\rm a~few})\times 10^6\,{\rm GeV}$ to make $f_{\rm R}\gtrsim 0.1\,{\rm Hz}$.
This lower bound on the inflation scale may be marginally allowed by the cosmic microwave background (CMB) observation~\cite{Planck:2018jri}. 
On the other hand, the ultimate-DECIGO can observe GWs with much lower inflation scale.

\begin{figure}[t]
    \centering
    \includegraphics[width=.45\textwidth ]{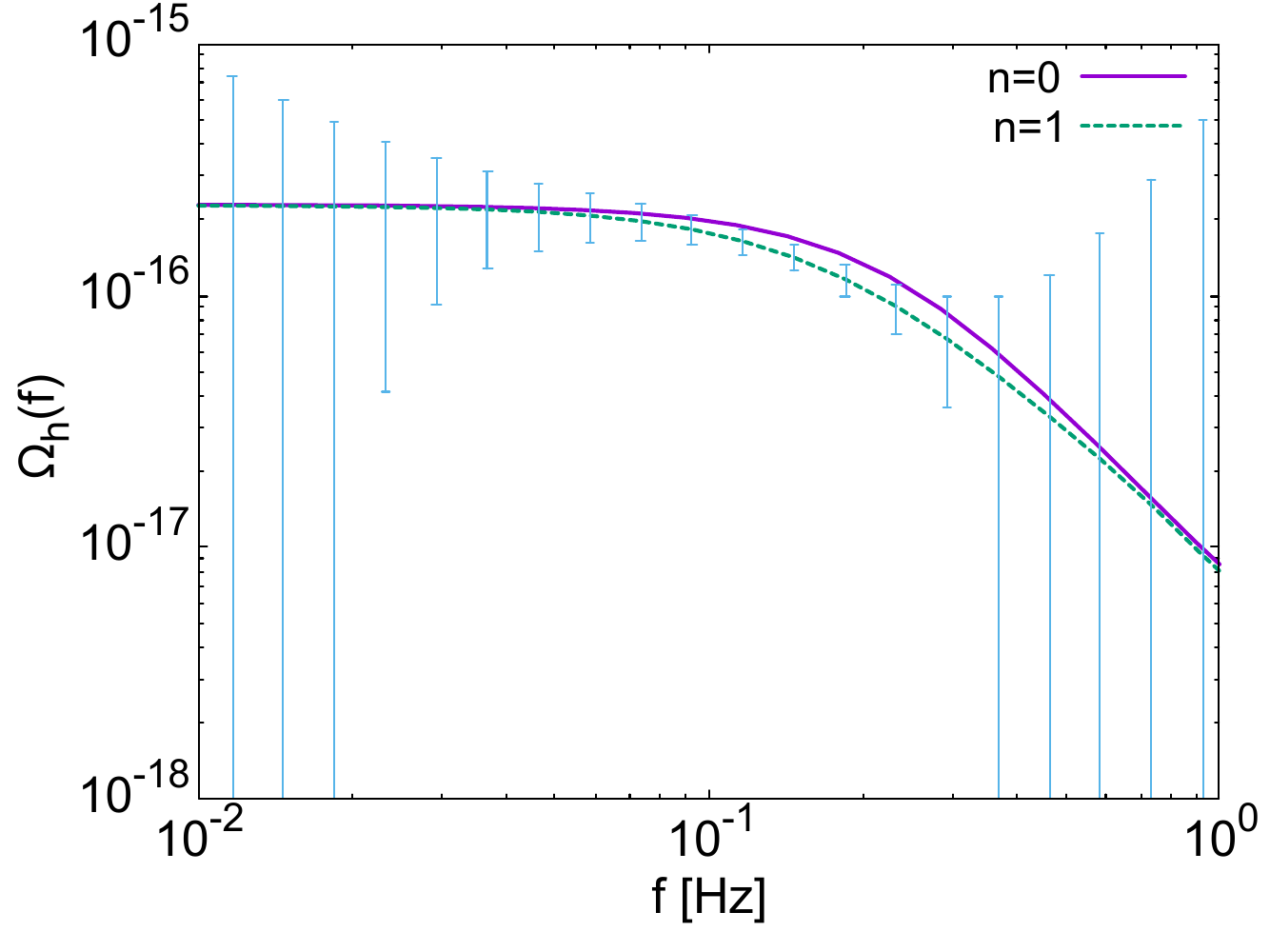}
    \hspace{5mm}
    \includegraphics[width=.45\textwidth ]{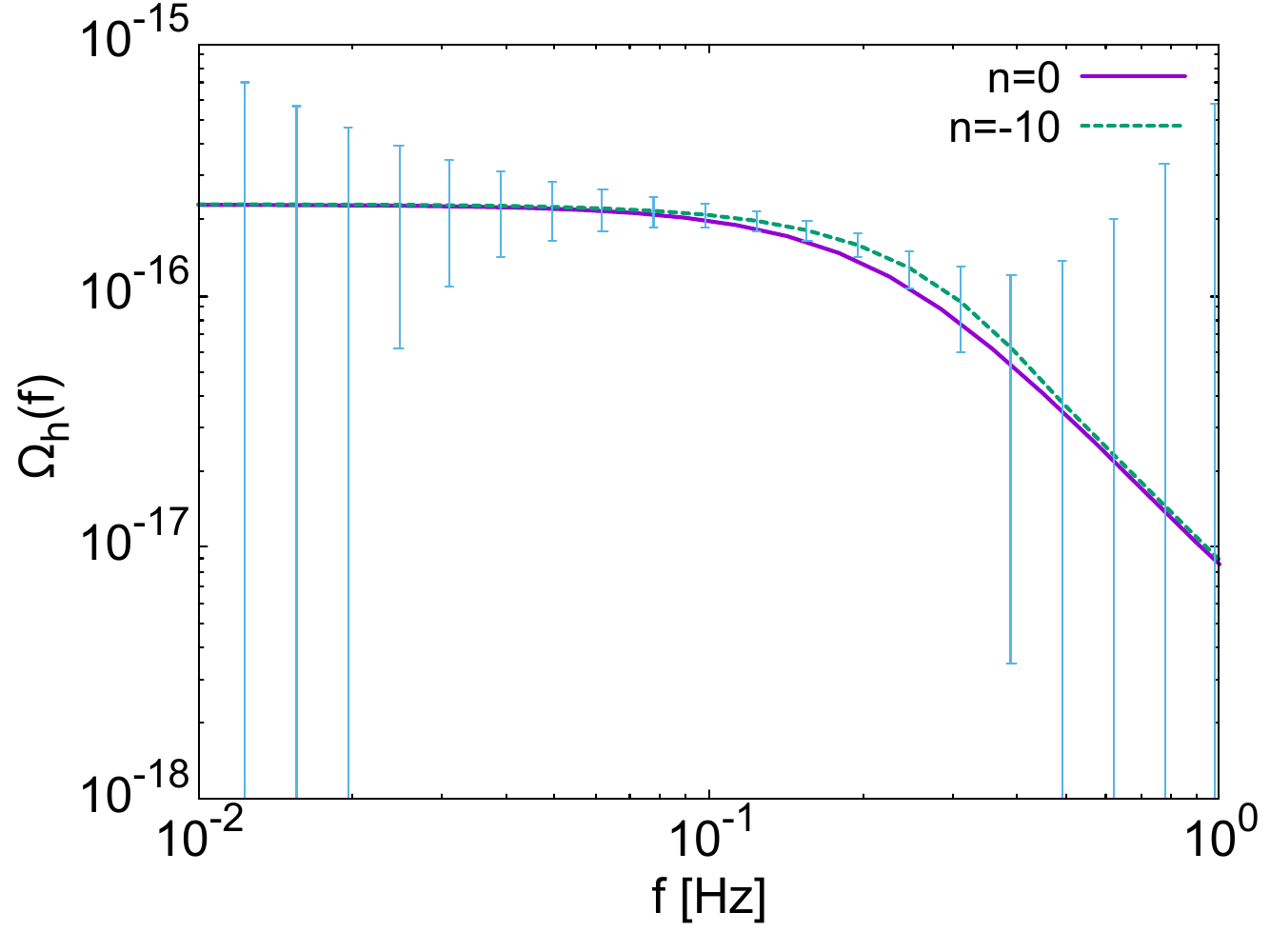}
    \caption{GW spectrum for $H_{\rm inf}=7\times 10^{13}\,{\rm GeV}$ and $T_{\rm R} = 6\times 10^6\,{\rm GeV}$ for $n=1$ (left) and $n=-10$ (right). For comparison, the case of $n=0$ is also shown. Error bars are based on the noise spectrum of FP-DECIGO with 10 years observation.}
    \label{fig:DECIGO}
\end{figure}

Fig.~\ref{fig:DECIGO} shows the GW spectrum for $H_{\rm inf}=7\times 10^{13}\,{\rm GeV}$ and $T_{\rm R} = 6\times 10^6\,{\rm GeV}$ for $n=1$ (left) and $n=-10$ (right). For comparison, the case of $n=0$ is also shown. Error bars are based on the noise spectrum of FP-DECIGO with 10 years observation.
It is seen that error bars at the most sensitive frequency ranges are comparable to the difference between two predictions.
We should note that, in order to truly distinguish these two lines, we need to observationally fix the both low and high frequency end of the spectrum. Otherwise, there remains a degeneracy between the change of $n$ and $T_{\rm R}$.
Thus it is fair to say that it is difficult to distinguish the dissipation model at in the FP-DECIGO adopted here, though a slight improvement on the sensitivity greatly improves this conclusion. 
 
\begin{figure}
    \centering
    \includegraphics[width=.45\textwidth ]{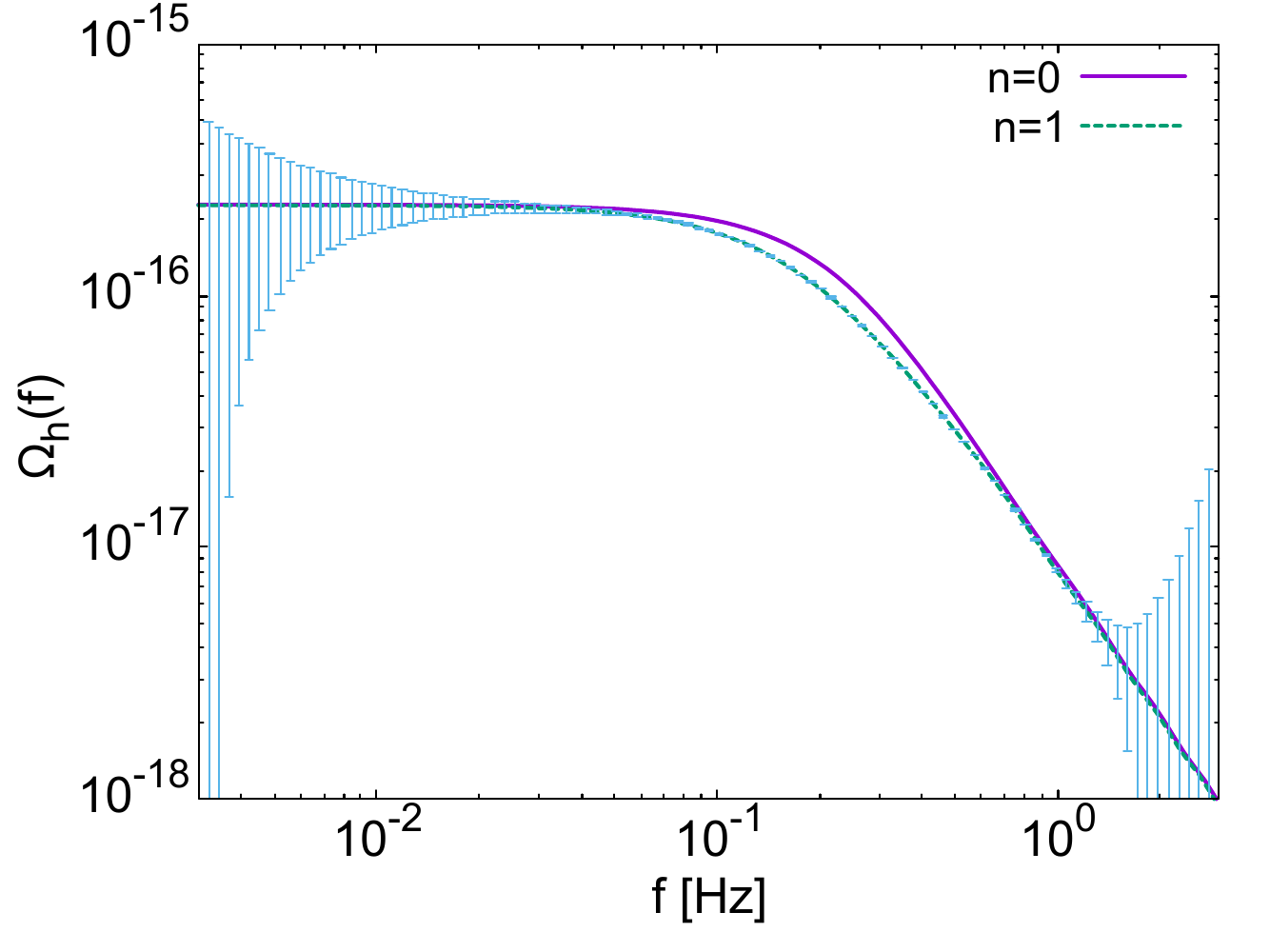}
    \hspace{5mm}
    \includegraphics[width=.45\textwidth ]{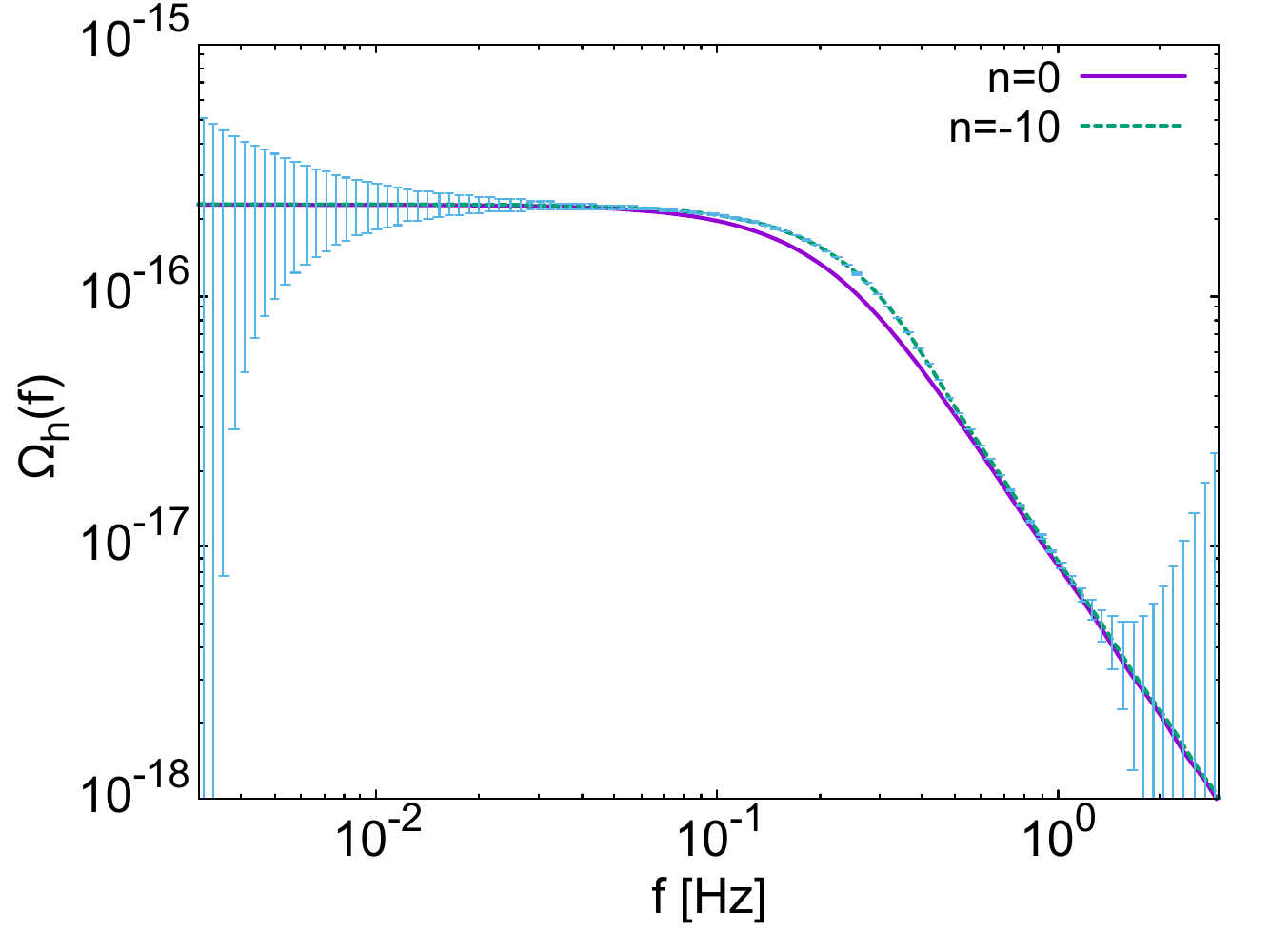}
    \caption{GW spectrum for $H_{\rm inf}=7\times 10^{13}\,{\rm GeV}$ and $T_{\rm R} = 6\times 10^6\,{\rm GeV}$ for $n=1$ (left) and $n=-10$ (right). For comparison, the case of $n=0$ is also shown. Error bars are based on the noise spectrum of ultimate-DECIGO with 10 years observation.}
    \label{fig:ultimate}
\end{figure}

\begin{figure}
    \centering
    \includegraphics[width=.45\textwidth ]{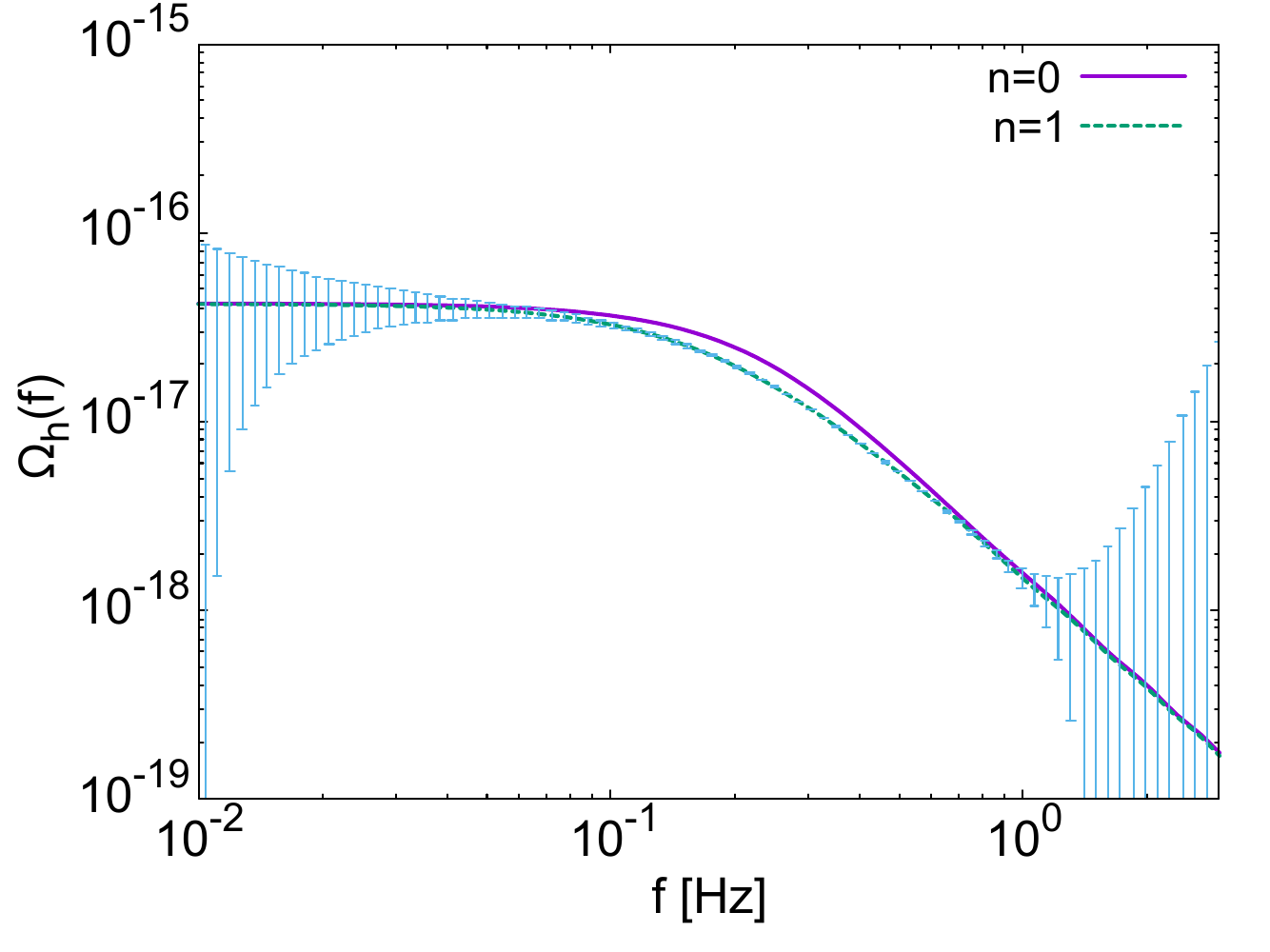}
    \hspace{5mm}
    \includegraphics[width=.45\textwidth ]{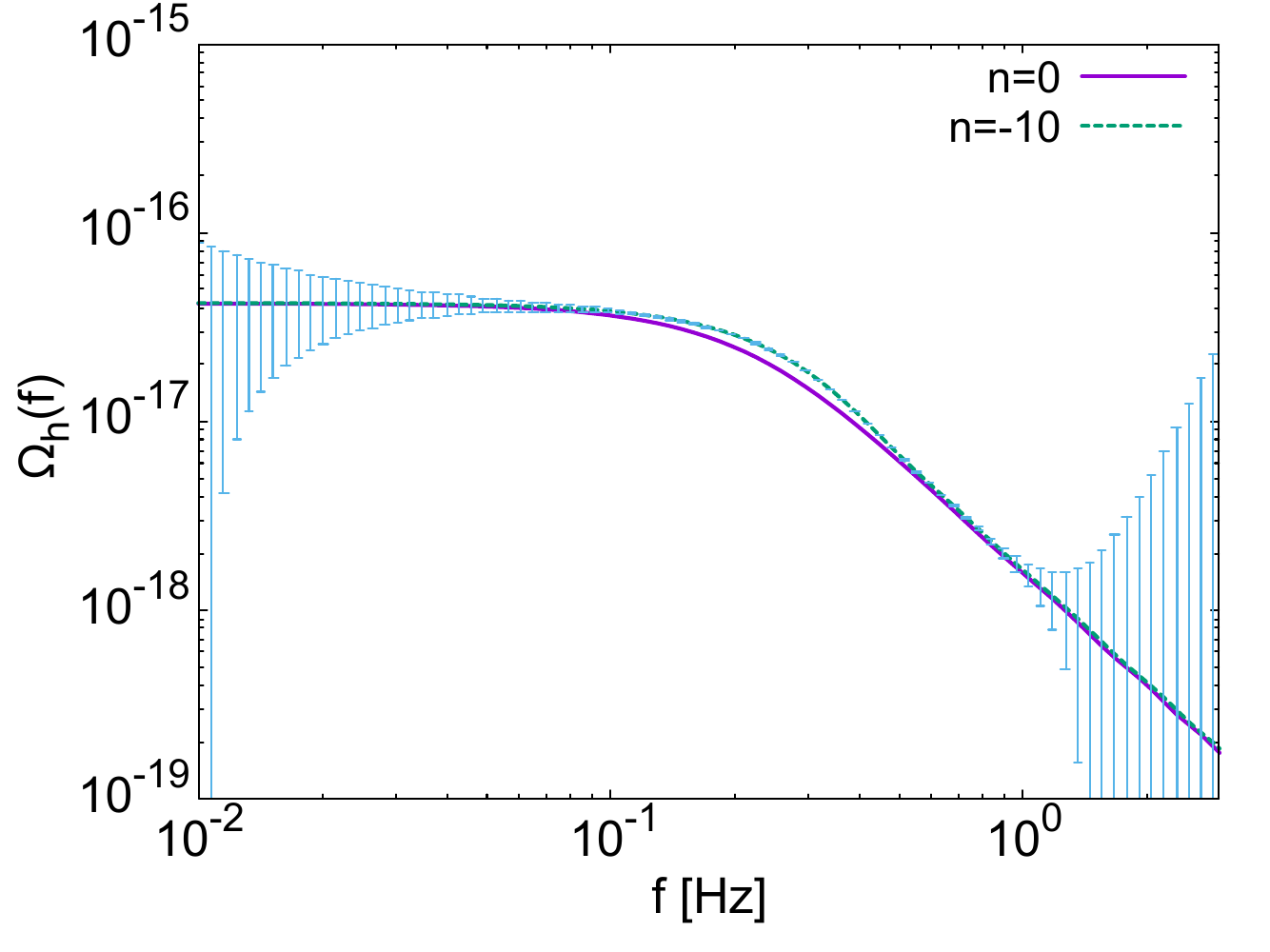}
    \caption{The same as Fig.~\ref{fig:ultimate}, but for $H_{\rm inf}=3\times 10^{13}\,{\rm GeV}$.}
    \label{fig:ultimate_H3e13}
\end{figure}

Fig.~\ref{fig:ultimate} shows the GW spectrum for $H_{\rm inf}=7\times 10^{13}\,{\rm GeV}$ and $T_{\rm R} = 6\times 10^6\,{\rm GeV}$ for $n=1$ (left) and $n=-10$ (right). Error bars are based on the noise spectrum of ultimate-DECIGO, which is adopted from Ref.~\cite{Kuroyanagi:2014qza}.
The same results but for $H_{\rm inf}=3\times 10^{13}\,{\rm GeV}$ are shown in Fig.~\ref{fig:ultimate_H3e13}.
Clearly the ultimate-DECIGO has a potential to distinguish the value of $n$, hence it not only determines the reheating temperature $T_{\rm R}$, but also how the reheating proceeds through thermal dissipation processes.
Of course, in this frequency range there should be huge foreground GWs from binaries of white dwarf or neutron stars/black holes~\cite{Farmer:2003pa,Regimbau:2011rp,Rosado:2011kv,KAGRA:2021kbb,Braglia:2022icu}.
The latter may be removed since their duty cycle is much smaller than unity, while the former rapidly falls off around 0.1~Hz.
In any case, this is just a demonstration that future GW observations with ultimate sensitivity have a potential to distinguish the physics of reheating and more realistic studies are beyond the scope of this paper.

\section{Conclusions}
\label{sec:conc}

Observations such as the CMB have already constrained inflationary models to some extent, but the reheating process remains entirely unknown. Various possibilities have been proposed for reheating, and if we could observationally distinguish among them, it would provide crucial information about the nature of the inflaton and other scalar fields. In this paper, we evaluated the spectrum of primordial GWs in the case where reheating is driven by thermal dissipation effects. We show that, while the overall spectrum is not drastically modified, its bending changes slightly. This difference can be detected by the FP-DECIGO or ultimate-DECIGO experiments, though some improvements of the sensitivity are required for FP-DECIGO.
Although we focused on thermal dissipation effects, it implies that other forms of time-dependent decay rate of the inflaton or any other scalar field may be imprinted in the GW spectrum.
For example, the decay rate may depend on some powers of the amplitude of a scalar field: $\Gamma \propto \phi(t)^n$. It may lead to similar signatures to those studied in the main text.
In any case, it shows that the observation of primordial GWs not only provides us with the information about reheating temperature but also the details of how physically the reheating proceeded.

\section*{Acknowledgment}

KN would like to thank S.~Kuroyanagi for discussion.
This work was supported by World Premier International Research Center Initiative (WPI), MEXT, Japan.
This work was also supported by JSPS KAKENHI (Grant Number 22K14044 [KM], 24K07010 [KN]).

\appendix

\section{Gravitational wave detection} \label{sec:noise}

In this Appendix we summarize the detector response to stochastic GWs and the noise spectrum expected at DECIGO.
We mostly follow Refs.~\cite{Maggiore:1999vm,Maggiore:2007ulw}.

\subsection{Detector response}

Let us consider correlation among several interferometers, labeled by $i$. Each detector output $N_i(t)$ is given by the sum of signal $s_i(t)$ and noise $n_i(t)$: $N_i(t) = s_i(t) + n_i(t)$.
The output is, for example, taken to be the relative modulation of the arm length in each interferometer caused by signal GWs or noises.
The noise spectrum is defined as
\begin{align}
	\left< n_i(t) n_j(t) \right> = \int_0^\infty df \frac{S_n(f)}{2} \delta_{ij}.
\end{align}
Its concrete form is given in the next subsection.

The signal $s_i(t)$ and the GW is related through the detector pattern function $F_i^\lambda(\hat\Omega)$ with $\hat\Omega$ being the arrival direction of the GW as, 
\begin{align}
	s_i(t) = \sum_\lambda \int df \int d\hat\Omega\,h_\lambda (f,\hat\Omega) F^\lambda_i(\hat\Omega) e^{-2\pi i f(t -\hat\Omega\cdot\vec x_i)},
\end{align}
where $\vec x_i$ is the position of each detector (one of the detector, $i=1$, can be assumed to be $\vec x_1=0$).
A concrete form is given by~\cite{Maggiore:1999vm}
\begin{align}
	&F^+_i(\theta,\varphi) = \frac{1}{2}\sin\alpha (1+\cos^2\theta) \sin(\alpha+2\varphi + 2\delta_i),\\
	&F^\times_i(\theta,\varphi) =-\sin\alpha \cos\theta \cos(\alpha+2\varphi + 2\delta_i),
\end{align}
where we have written $\hat \Omega=(\theta,\varphi)$, $\alpha$ is the angle between two arms in each interferometer (\textit{e.g.}, $\alpha=\pi/2$ for LIGO and $\alpha=\pi/3$ for DECIGO) and $\delta_i$ is the relative azimuthal angle of the detector: we can take $\delta_1=0$ and $\delta_2 = 2\pi/3$ for DECIGO.
We obtain
\begin{align}
	F_{ij}\equiv \int \frac{d\hat\Omega}{4\pi} \sum_\lambda F_i^\lambda(\hat\Omega) F_j^\lambda(\hat\Omega) = \frac{2}{5}\sin^2\alpha\cos(2\delta_{ij}),
	\label{Fij}
\end{align}
where $\delta_{ij}=\delta_i-\delta_j$.

Let us take the correlation of the two detector response as\footnote{
	Usually convolution is taken by involving the filter function $Q(t-t')$ with a choice of optimal filtering~\cite{Allen:1996vm}. Here we employ just a simple choice $Q(t-t')=\delta(t-t')$, but the resulting SNR does not change much as far as the frequency bin is taken small.
}
\begin{align}
	N_{12} = \int_{-t_{\rm obs}/2}^{t_{\rm obs}/2} dt\,N_1(t) N_2(t).
\end{align}
Since the noise is uncorrelated between different detectors, only the signal contributes to $\left< N_{12}\right>$. Thus the signal $S$ is given by
\begin{align}
	S = \frac{t_{\rm obs}}{2}\int_0^\infty df\,\Gamma(f) S_h(f),
\end{align}
where
\begin{align}
	\Gamma(f) \equiv \int \frac{d\hat\Omega}{4\pi} \left[ \sum_\lambda F^\lambda_1(\hat\Omega) F^\lambda_2(\hat\Omega) \right]e^{2\pi i f \hat\Omega\cdot \vec x_{2}}.
\end{align}
In the low frequency limit, it takes a simple form as $\Gamma(f) \simeq F_{12}$, given by (\ref{Fij}). On the other hand, the noise $N$ is given by
\begin{align}
	N = \sqrt{\left<N_{12}^2\right> - \left<N_{12}\right>^2} \simeq \left[ \frac{t_{\rm obs}}{8}\int_0^\infty df S_n^2(f)\right]^{1/2}.
\end{align}
We are interested in the SNR in a narrow frequency interval $\Delta f$ around $f$. Assuming that $S_h(f)$ and $S_n(f)$ do not change much within this frequency interval, we can define the SNR as
\begin{align}
	{\rm SNR}(f) \simeq \sqrt{2t_{\rm obs} \Delta f}F_{12}  \frac{S_h (f)}{S_n(f)}
	=  \sqrt{2t_{\rm obs} \Delta f} F_{12} \frac{\Omega_h(f)}{\Omega_n(f)}.
	\label{SNR_f12}
\end{align}
In the second equality we defined
\begin{align}
	\Omega_{n} (f) \equiv \frac{2\pi^2 f^3}{3H_0^2} S_{n}(f).
\end{align}
The remaining task is to find the noise spectrum $S_n(f)$.

Before doing so, let us comment on more specific case for the DECIGO~\cite{Kawamura:2020pcg}.
It consists of four interferometer clusters, each of which consists of three drag-free satellites in a triangle configuration. 
Among the four, two clusters are located at the same position on the Earth orbit and they are used for detecting stochastic GWs.
In each cluster labeled by $i=1,2$, there are three outputs from detectors at each satellite labeled by $X,Y,Z$.
Since an arm is shared by each interferometer, the noise in $X,Y$ and $Z$ are correlated.
For an equilateral triangle configuration, one can take a linear combination of the output so that the noise spectrum becomes diagonal~\cite{Ishikawa:2020hlo}:\footnote{
	The other linear combination leads to zero signal and hence we do not consider it here.
}
\begin{align}
	&A_i(t) \equiv \frac{1}{\sqrt 2}\left(N_{i,X}(t) - N_{i,Y}(t)\right) , \\
	&E_i(t) \equiv \frac{1}{\sqrt 6} \left(N_{i,X}(t) + N_{i,Y}(t) - 2 N_{i,Z}(t)\right).
\end{align}
The signal from the output $A_i$ is given by
\begin{align}
	\left<s^2_{A_i}(t) \right> =\frac{\sin^2\alpha}{5}\left( 1-\cos(2\delta_{XY})\right) \int_0^\infty df\,S_h(f).
\end{align}
The signal from the output $E$ is the same for DECIGO, where $\delta\equiv \delta_{XY} = \delta_{YZ} = \delta_{ZX} = 2\pi/3$.
Note the appearance of the factor $(1-\cos(2\delta)) = 3/2$ compared with a single detector. 
Now let us take a correlation between $A_1$ and $A_2$ to define a signal $S_A$:
\begin{align}
	S_A &= \int_{-t_{\rm obs}/2}^{t_{\rm obs}/2} dt\,s_{A_1} (t) s_{A_2}(t) \\
    &=t_{\rm obs}\frac{\sin^2\alpha}{5}\left( 1-\cos(2\delta)\right) \int_{0}^\infty df\,S_h(f).
\end{align}
We obtain the same expression for $S_E$, a signal from the correlation between $E_1$ and $E_2$.
Thus the total signal becomes factor 2 larger, while the noise obtains a factor of $\sqrt 2$.
Thus a factor $\sqrt{2}$ improvement for SNR is expected.
Taking these considerations into account, we replace $F_{12}$ in the expression of SNR (\ref{SNR_f12}) with $F$ defined as:
\begin{align}
	{\rm SNR}(f) =  \sqrt{2t_{\rm obs} \Delta f} \, F\,  \frac{\Omega_h(f)}{\Omega_n(f)},~~~~~~
	F \equiv  \frac{2\sqrt{2} \sin^2\alpha}{5}\left( 1-\cos(2\delta)\right).
\end{align}
For DECIGO, we obtain $F=9\sqrt{2}/20$.

\subsection{Noise spectrum}

We assume that $S_{n}(f)$ predominantly consists of quantum noises: shot noise and radiation pressure noise, 
\begin{align}
	S_{n}(f) = S_{\rm shot}(f) + S_{\rm rad}(f).
\end{align}
Other noises such as acceleration noise should be suppressed compared with these noises in order to achieve the desired sensitivity.
Below we give expressions for them~\cite{Kuroyanagi:2014qza,Ishikawa:2020hlo,Iwaguchi:2020cxa}.\footnote{
	Note that there is a typo in Eq.~(38) of Ref.~\cite{Kuroyanagi:2014qza}: a correct expression is $t_F = \sqrt{r_G^2-r_F^2}$, instead of $t_F = \sqrt{r_G^2-r_{Fm}^2}$. Also Eqs.~(34) and (35) of Ref.~\cite{Kuroyanagi:2014qza} should be understood as approximate results in the limit $r_G \sim 1$, $r_{F}\sim 1$ and $r_{E}=1$ after the replacement $\tilde P \to P$ in (34). (Note that $\tilde P\sim 4P$ in this limit.) The full expressions are found in Refs.~\cite{Ishikawa:2020hlo,Iwaguchi:2020cxa}.
}

Let us consider a Michelson interferometer with FP cavity, in which a laser light is reflected between the front mirror and end mirror, and their reflection/transmission coefficients are denoted by $r_i$ and $t_i$ ($i=1,2$ for the front and end mirror, respectively), satisfying $r_i^2+t_i^2=1$. 
The diffraction loss of the Gaussian beam is represented by $D_i$, which is given by
\begin{align}
	D_i^2 = 1- \exp\left( - \frac{2\pi R^2}{\lambda L}\right),
\end{align}
for optimized choice of Rayleigh length of the Gaussian beam, where $R$ is the mirror radius, $\lambda$ is the wavelength of the laser and $L$ is the arm length. We may define the effective reflection and transmission coefficients as $r_{{\rm eff},i} = D_i^2 r_i$, $t_{{\rm eff},i} = D_i^2 t_i$.
The cavity finesse is given by
\begin{align}
	\mathcal F = \frac{\pi \sqrt{ r_{{\rm eff},1} r_{{\rm eff},2} }}{1-r_{{\rm eff},1} r_{{\rm eff},2}}.
\end{align}
The full expression for shot noise and radiation pressure noise with these quantities are found in Ref.~\cite{Iwaguchi:2020cxa}.
Here we make use of some approximations: $r_2=1$ and $r_1 \sim 1$ and $D_i$ is also optimized so that $D_i \simeq 1$.
Then the shot noise and radiation pressure noise are given by
\begin{align}
	&S_{\rm shot}(f) =\left(\frac{1}{4\mathcal F L}\right)^2\frac{\pi\lambda}{\eta P} \left( 1 + \frac{f^2}{f_p^2} \right), \\
	&S_{\rm rad}(f) = \left(\frac{16\mathcal F}{LM (2\pi f)^2}\right)^2 \frac{P}{\pi\lambda}  \left( 1 + \frac{f^2}{f_p^2} \right)^{-1},
\end{align}
where $f_p = 1/(4\mathcal F L)$, $P$ is the laser power, $M$ is the mirror mass and $\eta$ is the efficiency of the photodetector, which is taken to be unity for simplicity.
Parameters we used for the FP-DECIGO and ultimate-DECIGO are summarized in Table~\ref{table}.
For the FP-DECIGO, they are taken from Ref.~\cite{Ishikawa:2020hlo,Iwaguchi:2020cxa}. On the other hand, the sensitivity for ultimate-DECIGO varies in the literature and there seems to be no consensus. It is not even clear whether the FP cavity setup is used or not.
Here we just take simple parameter sets with FP-DECIGO like setup to somehow mimic the sensitivity proposed so far.

\begin{table}
\begin{center}
\begin{tabular}{|c|c|c|c|c|c|} \hline
~ & $L$ & $\mathcal F$  & $P$  & $M$  & $\lambda$  \\ \hline
FP-DECIGO & $10^3\,{\rm km}$ & $10$  & $10\,{\rm W}$  & $100\,{\rm kg}$  & $515\,{\rm nm}$ \\ \hline 
ultimate-DECIGO & $5\times 10^4\,{\rm km}$ & $10$  & $10^3\,{\rm W}$  & $100\,{\rm kg}$  & $515\,{\rm nm}$ \\ \hline 
\end{tabular}
\caption{Parameters used for the sensitivity estimation of FP-DECIGO and ultimate-DECIGO.}
\label{table}
\end{center}
\end{table}

\bibliographystyle{utphys}
\bibliography{ref}

@article{Berges:2004yj,
    author = "Berges, Juergen",
    editor = "Bracco, Mirian and Chiapparini, Marcelo and Ferreira, Erasmo and Kodama, Takeshi",
    title = "{Introduction to nonequilibrium quantum field theory}",
    eprint = "hep-ph/0409233",
    archivePrefix = "arXiv",
    doi = "10.1063/1.1843591",
    journal = "AIP Conf. Proc.",
    volume = "739",
    number = "1",
    pages = "3--62",
    year = "2004"
}

@article{Kaneta:2025xuo,
    author = "Kaneta, Kunio and Takahashi, Tomo and Watanabe, Natsumi",
    title = "{Post-Reheating Inflaton Production as a Probe of Reheating Dynamics}",
    eprint = "2508.20402",
    archivePrefix = "arXiv",
    primaryClass = "hep-ph",
    month = "8",
    year = "2025"
}

@article{Fujita:2025zoa,
    author = "Fujita, Tomohiro and Mukaida, Kyohei and Tsuji, Tenta",
    title = "{Reheating after axion inflation}",
    eprint = "2503.01228",
    archivePrefix = "arXiv",
    primaryClass = "hep-ph",
    reportNumber = "KEK-TH-2692",
    doi = "10.1088/1475-7516/2025/07/002",
    journal = "JCAP",
    volume = "07",
    pages = "002",
    year = "2025"
}

@article{Glorioso:2016gsa,
    author = "Glorioso, Paolo and Liu, Hong",
    title = "{The second law of thermodynamics from symmetry and unitarity}",
    eprint = "1612.07705",
    archivePrefix = "arXiv",
    primaryClass = "hep-th",
    reportNumber = "MIT-CTP-4859",
    month = "12",
    year = "2016"
}

@article{Ema:2024hkj,
    author = "Ema, Yohei and Mukaida, Kyohei",
    title = "{Cutting rule for in-in correlators and cosmological collider}",
    eprint = "2409.07521",
    archivePrefix = "arXiv",
    primaryClass = "hep-th",
    reportNumber = "UMN-TH-4331/24, FTPI-MINN-24-18, KEK-TH-2653",
    doi = "10.1007/JHEP12(2024)194",
    journal = "JHEP",
    volume = "12",
    pages = "194",
    year = "2024"
}

@article{Liu:2018kfw,
    author = "Liu, Hong and Glorioso, Paolo",
    title = "{Lectures on non-equilibrium effective field theories and fluctuating hydrodynamics}",
    eprint = "1805.09331",
    archivePrefix = "arXiv",
    primaryClass = "hep-th",
    reportNumber = "MIT-CTP/5018; EFI-18-8, MIT-CTP-5018, EFI-18-8",
    doi = "10.22323/1.305.0008",
    journal = "PoS",
    volume = "TASI2017",
    pages = "008",
    year = "2018"
}

@article{Starobinsky:1979ty,
    author = "Starobinsky, Alexei A.",
    editor = "Khalatnikov, I. M. and Mineev, V. P.",
    title = "{Spectrum of relict gravitational radiation and the early state of the universe}",
    journal = "JETP Lett.",
    volume = "30",
    pages = "682--685",
    year = "1979"
}

@article{Starobinsky:1980te,
    author = "Starobinsky, Alexei A.",
    editor = "Khalatnikov, I. M. and Mineev, V. P.",
    title = "{A New Type of Isotropic Cosmological Models Without Singularity}",
    doi = "10.1016/0370-2693(80)90670-X",
    journal = "Phys. Lett. B",
    volume = "91",
    pages = "99--102",
    year = "1980"
}

@article{Guth:1980zm,
    author = "Guth, Alan H.",
    editor = "Fang, Li-Zhi and Ruffini, R.",
    title = "{The Inflationary Universe: A Possible Solution to the Horizon and Flatness Problems}",
    reportNumber = "SLAC-PUB-2576",
    doi = "10.1103/PhysRevD.23.347",
    journal = "Phys. Rev. D",
    volume = "23",
    pages = "347--356",
    year = "1981"
}

@article{Sato:1981qmu,
    author = "Sato, Katsuhiko",
    title = "{First-order phase transition of a vacuum and the expansion of the Universe}",
    doi = "10.1093/mnras/195.3.467",
    journal = "Mon. Not. Roy. Astron. Soc.",
    volume = "195",
    number = "3",
    pages = "467--479",
    year = "1981"
}

@article{Kazanas:1980tx,
    author = "Kazanas, D.",
    title = "{Dynamics of the Universe and Spontaneous Symmetry Breaking}",
    doi = "10.1086/183361",
    journal = "Astrophys. J. Lett.",
    volume = "241",
    pages = "L59--L63",
    year = "1980"
}

@article{Linde:1981mu,
    author = "Linde, Andrei D.",
    editor = "Fang, Li-Zhi and Ruffini, R.",
    title = "{A New Inflationary Universe Scenario: A Possible Solution of the Horizon, Flatness, Homogeneity, Isotropy and Primordial Monopole Problems}",
    reportNumber = "LEBEDEV-81-229",
    doi = "10.1016/0370-2693(82)91219-9",
    journal = "Phys. Lett. B",
    volume = "108",
    pages = "389--393",
    year = "1982"
}

@article{Albrecht:1982wi,
    author = "Albrecht, Andreas and Steinhardt, Paul J.",
    editor = "Fang, Li-Zhi and Ruffini, R.",
    title = "{Cosmology for Grand Unified Theories with Radiatively Induced Symmetry Breaking}",
    reportNumber = "UPR-0185T",
    doi = "10.1103/PhysRevLett.48.1220",
    journal = "Phys. Rev. Lett.",
    volume = "48",
    pages = "1220--1223",
    year = "1982"
}

@article{Allen:1987bk,
    author = "Allen, Bruce",
    title = "{The Stochastic Gravity Wave Background in Inflationary Universe Models}",
    reportNumber = "PRINT-88-0063 (TUFTS)",
    doi = "10.1103/PhysRevD.37.2078",
    journal = "Phys. Rev. D",
    volume = "37",
    pages = "2078",
    year = "1988"
}

@article{Turner:1990rc,
    author = "Turner, Michael S. and Wilczek, Frank",
    title = "{Relic gravitational waves and extended inflation}",
    reportNumber = "FERMILAB-PUB-90-178-A",
    doi = "10.1103/PhysRevLett.65.3080",
    journal = "Phys. Rev. Lett.",
    volume = "65",
    pages = "3080--3083",
    year = "1990"
}

@article{Turner:1993vb,
    author = "Turner, Michael S. and White, Martin J. and Lidsey, James E.",
    title = "{Tensor perturbations in inflationary models as a probe of cosmology}",
    eprint = "astro-ph/9306029",
    archivePrefix = "arXiv",
    reportNumber = "FERMILAB-PUB-93-069-A, CFPA-TH-93-19, CFPA-93-19",
    doi = "10.1103/PhysRevD.48.4613",
    journal = "Phys. Rev. D",
    volume = "48",
    pages = "4613--4622",
    year = "1993"
}

@article{Turner:1996ck,
    author = "Turner, Michael S.",
    title = "{Detectability of inflation produced gravitational waves}",
    eprint = "astro-ph/9607066",
    archivePrefix = "arXiv",
    reportNumber = "FERMILAB-PUB-96-169-A, FERMILAB-PUB-96-167-A",
    doi = "10.1103/PhysRevD.55.R435",
    journal = "Phys. Rev. D",
    volume = "55",
    pages = "R435--R439",
    year = "1997"
}

@article{Smith:2005mm,
    author = "Smith, Tristan L. and Kamionkowski, Marc and Cooray, Asantha",
    title = "{Direct detection of the inflationary gravitational wave background}",
    eprint = "astro-ph/0506422",
    archivePrefix = "arXiv",
    doi = "10.1103/PhysRevD.73.023504",
    journal = "Phys. Rev. D",
    volume = "73",
    pages = "023504",
    year = "2006"
}

@article{Seto:2003kc,
    author = "Seto, Naoki and Yokoyama, Jun'Ichi",
    title = "{Probing the equation of state of the early universe with a space laser interferometer}",
    eprint = "gr-qc/0305096",
    archivePrefix = "arXiv",
    reportNumber = "OU-TAP-206",
    doi = "10.1143/JPSJ.72.3082",
    journal = "J. Phys. Soc. Jap.",
    volume = "72",
    pages = "3082--3086",
    year = "2003"
}

@article{Tashiro:2003qp,
    author = "Tashiro, Hiroyuki and Chiba, Takeshi and Sasaki, Misao",
    title = "{Reheating after quintessential inflation and gravitational waves}",
    eprint = "gr-qc/0307068",
    archivePrefix = "arXiv",
    doi = "10.1088/0264-9381/21/7/004",
    journal = "Class. Quant. Grav.",
    volume = "21",
    pages = "1761--1772",
    year = "2004"
}

@article{Nakayama:2008ip,
    author = "Nakayama, Kazunori and Saito, Shun and Suwa, Yudai and Yokoyama, Jun'ichi",
    title = "{Space laser interferometers can determine the thermal history of the early Universe}",
    eprint = "0802.2452",
    archivePrefix = "arXiv",
    primaryClass = "hep-ph",
    reportNumber = "RESCEU-2-08, UTAP-594",
    doi = "10.1103/PhysRevD.77.124001",
    journal = "Phys. Rev. D",
    volume = "77",
    pages = "124001",
    year = "2008"
}

@article{Nakayama:2008wy,
    author = "Nakayama, Kazunori and Saito, Shun and Suwa, Yudai and Yokoyama, Jun'ichi",
    title = "{Probing reheating temperature of the universe with gravitational wave background}",
    eprint = "0804.1827",
    archivePrefix = "arXiv",
    primaryClass = "astro-ph",
    reportNumber = "RESCEU-7-08, UTAP-596",
    doi = "10.1088/1475-7516/2008/06/020",
    journal = "JCAP",
    volume = "06",
    pages = "020",
    year = "2008"
}

@article{Kuroyanagi:2008ye,
    author = "Kuroyanagi, Sachiko and Chiba, Takeshi and Sugiyama, Naoshi",
    title = "{Precision calculations of the gravitational wave background spectrum from inflation}",
    eprint = "0804.3249",
    archivePrefix = "arXiv",
    primaryClass = "astro-ph",
    doi = "10.1103/PhysRevD.79.103501",
    journal = "Phys. Rev. D",
    volume = "79",
    pages = "103501",
    year = "2009"
}

@article{Mukohyama:2009zs,
    author = "Mukohyama, Shinji and Nakayama, Kazunori and Takahashi, Fuminobu and Yokoyama, Shuichiro",
    title = "{Phenomenological Aspects of Horava-Lifshitz Cosmology}",
    eprint = "0905.0055",
    archivePrefix = "arXiv",
    primaryClass = "hep-th",
    reportNumber = "IPMU-09-0052",
    doi = "10.1016/j.physletb.2009.07.005",
    journal = "Phys. Lett. B",
    volume = "679",
    pages = "6--9",
    year = "2009"
}

@article{Nakayama:2009ce,
    author = "Nakayama, Kazunori and Yokoyama, Jun'ichi",
    title = "{Gravitational Wave Background and Non-Gaussianity as a Probe of the Curvaton Scenario}",
    eprint = "0910.0715",
    archivePrefix = "arXiv",
    primaryClass = "astro-ph.CO",
    reportNumber = "ICRR-REPORT-551, RESCEU-26-09",
    doi = "10.1088/1475-7516/2010/01/010",
    journal = "JCAP",
    volume = "01",
    pages = "010",
    year = "2010"
}

@article{Durrer:2011bi,
    author = "Durrer, Ruth and Hasenkamp, Jasper",
    title = "{Testing Superstring Theories with Gravitational Waves}",
    eprint = "1105.5283",
    archivePrefix = "arXiv",
    primaryClass = "gr-qc",
    doi = "10.1103/PhysRevD.84.064027",
    journal = "Phys. Rev. D",
    volume = "84",
    pages = "064027",
    year = "2011"
}

@article{Kuroyanagi:2011fy,
    author = "Kuroyanagi, Sachiko and Nakayama, Kazunori and Saito, Shun",
    title = "{Prospects for determination of thermal history after inflation with future gravitational wave detectors}",
    eprint = "1110.4169",
    archivePrefix = "arXiv",
    primaryClass = "astro-ph.CO",
    reportNumber = "ICRR-REPORT-597-2011-14, UT-11-35",
    doi = "10.1103/PhysRevD.84.123513",
    journal = "Phys. Rev. D",
    volume = "84",
    pages = "123513",
    year = "2011"
}

@article{Jinno:2014qka,
    author = "Jinno, Ryusuke and Moroi, Takeo and Takahashi, Tomo",
    title = "{Studying Inflation with Future Space-Based Gravitational Wave Detectors}",
    eprint = "1406.1666",
    archivePrefix = "arXiv",
    primaryClass = "astro-ph.CO",
    doi = "10.1088/1475-7516/2014/12/006",
    journal = "JCAP",
    volume = "12",
    pages = "006",
    year = "2014"
}

@article{Kuroyanagi:2014qza,
    author = "Kuroyanagi, Sachiko and Nakayama, Kazunori and Yokoyama, Jun'ichi",
    title = "{Prospects of determination of reheating temperature after inflation by DECIGO}",
    eprint = "1410.6618",
    archivePrefix = "arXiv",
    primaryClass = "astro-ph.CO",
    reportNumber = "RESCEU-44-14",
    doi = "10.1093/ptep/ptu176",
    journal = "PTEP",
    volume = "2015",
    number = "1",
    pages = "013E02",
    year = "2015"
}

@article{Jinno:2011sw,
    author = "Jinno, Ryusuke and Moroi, Takeo and Nakayama, Kazunori",
    title = "{Imprints of Cosmic Phase Transition in Inflationary Gravitational Waves}",
    eprint = "1112.0084",
    archivePrefix = "arXiv",
    primaryClass = "hep-ph",
    reportNumber = "UT-11-40",
    doi = "10.1016/j.physletb.2012.05.061",
    journal = "Phys. Lett. B",
    volume = "713",
    pages = "129--132",
    year = "2012"
}

@article{Weinberg:2003ur,
    author = "Weinberg, Steven",
    title = "{Damping of tensor modes in cosmology}",
    eprint = "astro-ph/0306304",
    archivePrefix = "arXiv",
    reportNumber = "UTTG-02-03",
    doi = "10.1103/PhysRevD.69.023503",
    journal = "Phys. Rev. D",
    volume = "69",
    pages = "023503",
    year = "2004"
}

@article{Dicus:2005rh,
    author = "Dicus, Duane A. and Repko, Wayne W.",
    title = "{Comment on damping of tensor modes in cosmology}",
    eprint = "astro-ph/0509096",
    archivePrefix = "arXiv",
    doi = "10.1103/PhysRevD.72.088302",
    journal = "Phys. Rev. D",
    volume = "72",
    pages = "088302",
    year = "2005"
}

@article{Boyle:2005se,
    author = "Boyle, Latham A. and Steinhardt, Paul J.",
    title = "{Probing the early universe with inflationary gravitational waves}",
    eprint = "astro-ph/0512014",
    archivePrefix = "arXiv",
    doi = "10.1103/PhysRevD.77.063504",
    journal = "Phys. Rev. D",
    volume = "77",
    pages = "063504",
    year = "2008"
}

@article{Watanabe:2006qe,
    author = "Watanabe, Yuki and Komatsu, Eiichiro",
    title = "{Improved Calculation of the Primordial Gravitational Wave Spectrum in the Standard Model}",
    eprint = "astro-ph/0604176",
    archivePrefix = "arXiv",
    doi = "10.1103/PhysRevD.73.123515",
    journal = "Phys. Rev. D",
    volume = "73",
    pages = "123515",
    year = "2006"
}

@article{Jinno:2012xb,
    author = "Jinno, Ryusuke and Moroi, Takeo and Nakayama, Kazunori",
    title = "{Probing dark radiation with inflationary gravitational waves}",
    eprint = "1208.0184",
    archivePrefix = "arXiv",
    primaryClass = "astro-ph.CO",
    reportNumber = "UT-12-26",
    doi = "10.1103/PhysRevD.86.123502",
    journal = "Phys. Rev. D",
    volume = "86",
    pages = "123502",
    year = "2012"
}

@article{Jinno:2013xqa,
    author = "Jinno, Ryusuke and Moroi, Takeo and Nakayama, Kazunori",
    title = "{Inflationary Gravitational Waves and the Evolution of the Early Universe}",
    eprint = "1307.3010",
    archivePrefix = "arXiv",
    primaryClass = "hep-ph",
    reportNumber = "UT-13-27",
    doi = "10.1088/1475-7516/2014/01/040",
    journal = "JCAP",
    volume = "01",
    pages = "040",
    year = "2014"
}

@article{Saikawa:2018rcs,
    author = "Saikawa, Ken'ichi and Shirai, Satoshi",
    title = "{Primordial gravitational waves, precisely: The role of thermodynamics in the Standard Model}",
    eprint = "1803.01038",
    archivePrefix = "arXiv",
    primaryClass = "hep-ph",
    reportNumber = "IPMU18-0037, MPP-2018-19",
    doi = "10.1088/1475-7516/2018/05/035",
    journal = "JCAP",
    volume = "05",
    pages = "035",
    year = "2018"
}

@article{Ringwald:2020vei,
    author = "Ringwald, Andreas and Saikawa, Ken'ichi and Tamarit, Carlos",
    title = "{Primordial gravitational waves in a minimal model of particle physics and cosmology}",
    eprint = "2009.02050",
    archivePrefix = "arXiv",
    primaryClass = "hep-ph",
    reportNumber = "DESY 20-135, DESY-20-135, KANAZAWA-20-06, TUM-HEP-1279-20",
    doi = "10.1088/1475-7516/2021/02/046",
    journal = "JCAP",
    volume = "02",
    pages = "046",
    year = "2021"
}

@article{Seto:2001qf,
    author = "Seto, Naoki and Kawamura, Seiji and Nakamura, Takashi",
    title = "{Possibility of direct measurement of the acceleration of the universe using 0.1-Hz band laser interferometer gravitational wave antenna in space}",
    eprint = "astro-ph/0108011",
    archivePrefix = "arXiv",
    doi = "10.1103/PhysRevLett.87.221103",
    journal = "Phys. Rev. Lett.",
    volume = "87",
    pages = "221103",
    year = "2001"
}

@article{Kudoh:2005as,
    author = "Kudoh, Hideaki and Taruya, Atsushi and Hiramatsu, Takashi and Himemoto, Yoshiaki",
    title = "{Detecting a gravitational-wave background with next-generation space interferometers}",
    eprint = "gr-qc/0511145",
    archivePrefix = "arXiv",
    reportNumber = "UTAP-544, RESCEU-37-05",
    doi = "10.1103/PhysRevD.73.064006",
    journal = "Phys. Rev. D",
    volume = "73",
    pages = "064006",
    year = "2006"
}

@inproceedings{Allen:1996vm,
    author = "Allen, Bruce",
    title = "{The Stochastic gravity wave background: Sources and detection}",
    booktitle = "{Les Houches School of Physics: Astrophysical Sources of Gravitational Radiation}",
    eprint = "gr-qc/9604033",
    archivePrefix = "arXiv",
    reportNumber = "WISC-MILW-96-TH-22",
    pages = "373--417",
    month = "4",
    year = "1996"
}

@article{Allen:1997ad,
    author = "Allen, Bruce and Romano, Joseph D.",
    title = "{Detecting a stochastic background of gravitational radiation: Signal processing strategies and sensitivities}",
    eprint = "gr-qc/9710117",
    archivePrefix = "arXiv",
    reportNumber = "WISC-MILW-97-TH-14",
    doi = "10.1103/PhysRevD.59.102001",
    journal = "Phys. Rev. D",
    volume = "59",
    pages = "102001",
    year = "1999"
}

@article{Maggiore:1999vm,
    author = "Maggiore, Michele",
    title = "{Gravitational wave experiments and early universe cosmology}",
    eprint = "gr-qc/9909001",
    archivePrefix = "arXiv",
    reportNumber = "IFUP-TH-20-99",
    doi = "10.1016/S0370-1573(99)00102-7",
    journal = "Phys. Rept.",
    volume = "331",
    pages = "283--367",
    year = "2000"
}

@book{Maggiore:2007ulw,
    author = "Maggiore, Michele",
    title = "{Gravitational Waves. Vol. 1: Theory and Experiments}",
    doi = "10.1093/acprof:oso/9780198570745.001.0001",
    isbn = "978-0-19-171766-6, 978-0-19-852074-0",
    publisher = "Oxford University Press",
    year = "2007"
}

@article{Farmer:2003pa,
    author = "Farmer, Alison J. and Phinney, E. Sterl",
    title = "{The gravitational wave background from cosmological compact binaries}",
    eprint = "astro-ph/0304393",
    archivePrefix = "arXiv",
    doi = "10.1111/j.1365-2966.2003.07176.x",
    journal = "Mon. Not. Roy. Astron. Soc.",
    volume = "346",
    pages = "1197",
    year = "2003"
}

@article{Regimbau:2011rp,
    author = "Regimbau, Tania",
    title = "{The astrophysical gravitational wave stochastic background}",
    eprint = "1101.2762",
    archivePrefix = "arXiv",
    primaryClass = "astro-ph.CO",
    doi = "10.1088/1674-4527/11/4/001",
    journal = "Res. Astron. Astrophys.",
    volume = "11",
    pages = "369--390",
    year = "2011"
}

@article{Rosado:2011kv,
    author = "Rosado, Pablo A.",
    title = "{Gravitational wave background from binary systems}",
    eprint = "1106.5795",
    archivePrefix = "arXiv",
    primaryClass = "gr-qc",
    doi = "10.1103/PhysRevD.84.084004",
    journal = "Phys. Rev. D",
    volume = "84",
    pages = "084004",
    year = "2011"
}

@article{KAGRA:2021kbb,
    author = "Abbott, R. and others",
    collaboration = "KAGRA, Virgo, LIGO Scientific",
    title = "{Upper limits on the isotropic gravitational-wave background from Advanced LIGO and Advanced Virgo{\textquoteright}s third observing run}",
    eprint = "2101.12130",
    archivePrefix = "arXiv",
    primaryClass = "gr-qc",
    reportNumber = "LIGO-DCC-P2000314",
    doi = "10.1103/PhysRevD.104.022004",
    journal = "Phys. Rev. D",
    volume = "104",
    number = "2",
    pages = "022004",
    year = "2021"
}

@article{Braglia:2022icu,
    author = "Braglia, Matteo and Garcia-Bellido, Juan and Kuroyanagi, Sachiko",
    title = "{Tracking the origin of black holes with the stochastic gravitational wave background popcorn signal}",
    eprint = "2201.13414",
    archivePrefix = "arXiv",
    primaryClass = "astro-ph.CO",
    reportNumber = "IFT-UAM/CSIC-22-6",
    doi = "10.1093/mnras/stad082",
    journal = "Mon. Not. Roy. Astron. Soc.",
    volume = "519",
    number = "4",
    pages = "6008--6019",
    year = "2023"
}

@article{Ishikawa:2020hlo,
    author = "Ishikawa, Tomohiro and others",
    title = "{Improvement of the target sensitivity in DECIGO by optimizing its parameters for quantum noise including the effect of diffraction loss}",
    eprint = "2012.11859",
    archivePrefix = "arXiv",
    primaryClass = "gr-qc",
    doi = "10.3390/galaxies9010014",
    journal = "Galaxies",
    volume = "9",
    number = "1",
    pages = "14",
    year = "2021"
}

@article{Iwaguchi:2020cxa,
    author = "Iwaguchi, Shoki. and others",
    title = "{Quantum noise in a Fabry-Perot interferometer including the influence of diffraction loss of light}",
    eprint = "2012.12530",
    archivePrefix = "arXiv",
    primaryClass = "gr-qc",
    doi = "10.3390/galaxies9010009",
    journal = "Galaxies",
    volume = "9",
    number = "1",
    pages = "9",
    year = "2021"
}

@article{Yokoyama:2004pf,
    author = "Yokoyama, Jun'ichi",
    title = "{Fate of oscillating scalar fields in the thermal bath and their cosmological implications}",
    eprint = "hep-ph/0406072",
    archivePrefix = "arXiv",
    reportNumber = "OU-TAP-232",
    doi = "10.1103/PhysRevD.70.103511",
    journal = "Phys. Rev. D",
    volume = "70",
    pages = "103511",
    year = "2004"
}

@article{Yokoyama:2005dv,
    author = "Yokoyama, Jun'ichi",
    title = "{Can oscillating scalar fields decay into particles with a large thermal mass?}",
    eprint = "hep-ph/0510091",
    archivePrefix = "arXiv",
    reportNumber = "RESCEU-15-05",
    doi = "10.1016/j.physletb.2006.02.039",
    journal = "Phys. Lett. B",
    volume = "635",
    pages = "66--71",
    year = "2006"
}

@article{Drewes:2010pf,
    author = "Drewes, Marco",
    title = "{On the Role of Quasiparticles and thermal Masses in Nonequilibrium Processes in a Plasma}",
    eprint = "1012.5380",
    archivePrefix = "arXiv",
    primaryClass = "hep-th",
    month = "12",
    year = "2010"
}

@article{Bastero-Gil:2010dgy,
    author = "Bastero-Gil, Mar and Berera, Arjun and Ramos, Rudnei O.",
    title = "{Dissipation coefficients from scalar and fermion quantum field interactions}",
    eprint = "1008.1929",
    archivePrefix = "arXiv",
    primaryClass = "hep-ph",
    doi = "10.1088/1475-7516/2011/09/033",
    journal = "JCAP",
    volume = "09",
    pages = "033",
    year = "2011"
}

@article{Mukaida:2012qn,
    author = "Mukaida, Kyohei and Nakayama, Kazunori",
    title = "{Dynamics of oscillating scalar field in thermal environment}",
    eprint = "1208.3399",
    archivePrefix = "arXiv",
    primaryClass = "hep-ph",
    reportNumber = "UT-12-28",
    doi = "10.1088/1475-7516/2013/01/017",
    journal = "JCAP",
    volume = "01",
    pages = "017",
    year = "2013"
}

@article{Mukaida:2012bz,
    author = "Mukaida, Kyohei and Nakayama, Kazunori",
    title = "{Dissipative Effects on Reheating after Inflation}",
    eprint = "1212.4985",
    archivePrefix = "arXiv",
    primaryClass = "hep-ph",
    reportNumber = "UT-12-45",
    doi = "10.1088/1475-7516/2013/03/002",
    journal = "JCAP",
    volume = "03",
    pages = "002",
    year = "2013"
}

@article{Mukaida:2013xxa,
    author = "Mukaida, Kyohei and Nakayama, Kazunori and Takimoto, Masahiro",
    title = "{Fate of $Z_2$ Symmetric Scalar Field}",
    eprint = "1308.4394",
    archivePrefix = "arXiv",
    primaryClass = "hep-ph",
    reportNumber = "UT-13-30",
    doi = "10.1007/JHEP12(2013)053",
    journal = "JHEP",
    volume = "12",
    pages = "053",
    year = "2013"
}

@article{Drewes:2013iaa,
    author = "Drewes, Marco and Kang, Jin U",
    title = "{The Kinematics of Cosmic Reheating}",
    eprint = "1305.0267",
    archivePrefix = "arXiv",
    primaryClass = "hep-ph",
    reportNumber = "TUM-HEP-886-13, CAS-KITPC-ITP-367",
    doi = "10.1016/j.nuclphysb.2013.07.009",
    journal = "Nucl. Phys. B",
    volume = "875",
    pages = "315--350",
    year = "2013",
    note = "[Erratum: Nucl.Phys.B 888, 284--286 (2014)]"
}

@article{Mukaida:2014yia,
    author = "Mukaida, Kyohei and Nakayama, Kazunori and Takimoto, Masahiro",
    title = "{Curvaton Dynamics Revisited}",
    eprint = "1401.5821",
    archivePrefix = "arXiv",
    primaryClass = "hep-ph",
    reportNumber = "UT-14-02",
    doi = "10.1088/1475-7516/2014/06/013",
    journal = "JCAP",
    volume = "06",
    pages = "013",
    year = "2014"
}

@article{Mukaida:2014kpa,
    author = "Mukaida, Kyohei and Nakayama, Kazunori",
    title = "{Dark Matter Chaotic Inflation in Light of BICEP2}",
    eprint = "1404.1880",
    archivePrefix = "arXiv",
    primaryClass = "hep-ph",
    reportNumber = "UT-14-15",
    doi = "10.1088/1475-7516/2014/08/062",
    journal = "JCAP",
    volume = "08",
    pages = "062",
    year = "2014"
}

@article{Tanin:2017bzm,
    author = "Tanin, Erwin H. and Stewart, Ewan D.",
    title = "{Damping of an oscillating scalar field indirectly coupled to a thermal bath}",
    eprint = "1708.04865",
    archivePrefix = "arXiv",
    primaryClass = "hep-ph",
    doi = "10.1088/1475-7516/2017/11/019",
    journal = "JCAP",
    volume = "11",
    pages = "019",
    year = "2017"
}

@article{Ai:2021gtg,
    author = "Ai, Wen-Yuan and Drewes, Marco and Glavan, Dra{\v{z}}en and Hajer, Jan",
    title = "{Oscillating scalar dissipating in a medium}",
    eprint = "2108.00254",
    archivePrefix = "arXiv",
    primaryClass = "hep-ph",
    reportNumber = "CP3-21-48",
    doi = "10.1007/JHEP11(2021)160",
    journal = "JHEP",
    volume = "11",
    pages = "160",
    year = "2021"
}

@article{Ai:2023ahr,
    author = "Ai, Wen-Yuan and Wang, Zi-Liang",
    title = "{Fate of oscillating homogeneous {\ensuremath{\mathbb{Z}}}$_{2}$-symmetric scalar condensates in the early Universe}",
    eprint = "2307.14811",
    archivePrefix = "arXiv",
    primaryClass = "hep-ph",
    reportNumber = "KCL-PH-TH/2023-41",
    doi = "10.1088/1475-7516/2024/06/075",
    journal = "JCAP",
    volume = "06",
    pages = "075",
    year = "2024"
}

@article{Planck:2018jri,
    author = "Akrami, Y. and others",
    collaboration = "Planck",
    title = "{Planck 2018 results. X. Constraints on inflation}",
    eprint = "1807.06211",
    archivePrefix = "arXiv",
    primaryClass = "astro-ph.CO",
    doi = "10.1051/0004-6361/201833887",
    journal = "Astron. Astrophys.",
    volume = "641",
    pages = "A10",
    year = "2020"
}

@article{Kawamura:2020pcg,
    author = "Kawamura, Seiji and others",
    title = "{Current status of space gravitational wave antenna DECIGO and B-DECIGO}",
    eprint = "2006.13545",
    archivePrefix = "arXiv",
    primaryClass = "gr-qc",
    doi = "10.1093/ptep/ptab019",
    journal = "PTEP",
    volume = "2021",
    number = "5",
    pages = "05A105",
    year = "2021"
}

\end{document}